%% file: main.tex
\documentclass[fleqn,10pt]{wlscirep}
\usepackage[utf8]{inputenc}
\usepackage[T1]{fontenc}
\usepackage{makecell}
\usepackage{tikz}
\usetikzlibrary{shapes.geometric}
\usetikzlibrary{shapes.arrows}
\usetikzlibrary{calc}
\usetikzlibrary{positioning}
\usepackage{array}
\usetikzlibrary{trees}
\usepackage[]{graphicx}
\usepackage{diagbox}
\usepackage{chronology}
\usepackage{booktabs}
\usepackage{multirow}

\usepackage{url}
\usepackage{amssymb}
\usepackage{subfigure}
\usepackage{tabularx}
\usepackage{dcolumn}
\usepackage{bm}
\usepackage{breqn}
\usepackage[export]{adjustbox}
\usepackage{colortbl}
\usepackage[braket, qm]{qcircuit} 
\usepackage[mathlines]{lineno}

\usepackage{xcolor}
\usepackage[]{hyperref}
\hypersetup{colorlinks=true, citecolor=blue, linkcolor = red,}

\newcommand{\NATURE}[1]{}
\newcommand{\ARXIV}[1]{#1}

\input{_commands}

\title{A Higher Radix Architecture for Quantum Carry-lookahead Adder}

\author[*,1]{Siyi Wang} 
\author[*,2]{Anubhab Baksi}
\author[*,3]{Anupam Chattopadhyay}

\affil[*]{School of Computer Science and Engineering, Nanyang~Technological~University, Singapore, 639798}

\affil[1]{siyi002@e.ntu.edu.sg}
\affil[2]{anubhab.baksi@ntu.edu.sg}
\affil[3]{anupam@ntu.edu.sg}

\begin{abstract}
\input{document/0-abstract}
\end{abstract}
\begin{document}

\flushbottom
\maketitle
%
%
\thispagestyle{empty}


\NATURE{\section*{Introduction}}\ARXIV{\section{Introduction}}
\input{document/1-introduction}

\NATURE{\section*{Previous Works}}\ARXIV{\section{Previous Works}}\label{s:adder-prev}
\input{document/2-0-previous-works}

\NATURE{\section*{Method}}\ARXIV{\section{Method}}\label{s:adder-stru}
\input{document/3-0-description}

\NATURE{\subsection*{Basic Component:  C$_n$NOT}}\ARXIV{\subsection{Basic Component:  C$_n$NOT}}
\input{document/3-1-basic-component}
\NATURE{\subsection*{Step 1: Higher Radix Layer}}\ARXIV{\subsection{Step 1: Higher Radix Layer}}
\input{document/3-2-higher-radix-layer}

\NATURE{\subsection*{Step 2: Carry Path}}\ARXIV{\subsection{Step 2: Carry Path}}\label{Carry path}
\input{document/3-3-carry-path}

\NATURE{\subsection*{Step 3: Sum Path}}\ARXIV{\subsection{Step 3: Sum Path}}\label{Sum path}
\input{document/3-4-sum-path}
\NATURE{\subsection*{Overall Structure of Quantum Higher Radix Adder}}\ARXIV{\subsection{Overall Structure of Quantum Higher Radix Adder}}\label{Overall}
\input{document/3-5-Overall}

\NATURE{\section*{Results and Discussions}}\ARXIV{\section{Results and Discussions}} \label{s:adder-resu}



\begin{itemize}
\item \textbf{Experiment 1: The effect of radix.}\\
\input{document/4-1-Experiment-1}
\item \textbf{Selection of Best Radix.}\\
\input{document/4-2-Selection}
\item \textbf{Experiment 2: Comparison with well-known quantum adders.}\\
\input{document/4-3-Experiment-2}
\item \textbf{Connecting with existing quantum adders.}\\
\input{document/4-4-Connecting}

\end{itemize}


\NATURE{\section*{Conclusion}}\ARXIV{\section{Conclusion}} \label{s:adder-conc}
\input{document/5-conclusion}

\NATURE{
\section*{Data Availability}
The data that support the findings of this study are available from the corresponding author upon reasonable request.
}

\NATURE{
\section*{Code Availability}
}
The relevant code will be available as a public repository (\url{https://github.com/Siyi-06/Quantum_higher_radix_adder}).

\bibliography{main}



\NATURE{
\section*{Author contributions statement}
S.W. and A.C. conceived the circuit structure. S.W. wrote the main manuscript text. S.W. and A.B. polished the draft and drew the diagrams as well as tables. All authors reviewed the manuscript.
}

\NATURE{
\section*{Competing interests}
All authors declare no financial or non-financial competing interests. 
}

\NATURE{
\section*{Additional information}
Correspondence and requests for materials should be addressed to S.W.
}


\appendix
\section{Worked-out Examples}\label{appendix6-adder}\input{document/6-appendix}
\section{Derivation Details of Cost Formulae}\label{appendix7-adder} \input{document/7-appendix}
\section{Optimum Radix}\label{best-r}\input{document/8-appendix}
\NATURE{\input{document/10-appendix}}
\end{document}

%% file: _commands.tex
\usepackage{xspace}

\newcommand{\secn}{Section\xspace}

\newcommand{\attherate}{\textcolor{teal}{$\diamond$}\xspace}
\newcommand{\exclamation}{\textcolor{teal}{$\star$}\xspace}
\newcommand{\mysignal}{\textcolor{teal}{$\bullet$}\xspace}

\newcommand{\RNum}[1]{\uppercase\expandafter{\romannumeral #1\relax}}

%% file: document/0-abstract.tex
In this paper, we propose an efficient quantum carry-lookahead adder based on the higher radix structure. For the addition of two $n$-bit numbers, our adder uses $O(n)-O(\frac{n}{r})$ qubits and $O(n)+O(\frac{n}{r})$ T gates to get the correct answer in T-depth $O(r)+O(\log{\frac{n}{r}})$, where $r$ is the radix.

Quantum carry-lookahead adder has already attracted some attention because of its low T-depth. Our work further reduces the overall cost by introducing a higher radix layer. By analyzing the performance in T-depth, T-count, and qubit count, it is shown that the proposed adder is superior to existing quantum carry-lookahead adders. Even compared to the Draper out-of-place adder which is very compact and efficient, our adder is still better in terms of T-count.

%% file: document/1-introduction.tex
As the field of quantum computing has been gaining momentum over the last few years, the need for optimizing quantum circuits is also growing. Quantum adders are one of the most important basic components of quantum computing circuits. The continuous development of quantum adders has not only improved the efficiency of small basic quantum computing circuits such as multiplication circuits but also has a significant effect on some prominent large quantum circuits. 
On the one hand, an efficient quantum adder can increase the speed of quantum addition and multiplication operations and reduce the cost of the required resources. 
On the other hand, quantum adders are widely used in Shor's algorithm\cite{Shor} which plays an important role in the field of public key cryptography. 
Therefore, an efficient quantum adder not only has significant financial benefits but also makes a crucial contribution to the development of quantum computing.

Even though the adders are highly analyzed in classical computing, we observe that the niche is still not properly studied in the quantum paradigm.
In the field of quantum adders, the quantum ripple carry adder (RCA)\cite{VBE, Cuccaro} was first proposed. However, the T-depth of quantum RCAs increases linearly with the number of input qubits, which means they need a long time to perform the operation.
Then some quantum carry look-ahead adder (CLA) designs such as Draper's logarithmic adder\cite{Draper08}, whose T-depth increases logarithmically with the number of input qubits, have been proposed to get further efficiency gains.
In the field of the classical adder, Gurkaynak\cite{classic_higher_radix} et al. found that increasing the radix of the CLA can effectively decrease the computation time of classical CLAs. 
However, the problem of quantum CLA implementation has not been explored yet, to the best of our finding.

The objective of this work is to explore the potential of higher radix strategy in improving the performance of quantum arithmetic circuits.
As we will see later, high fan-out is a challenge for quantum adder.
Using the idea of separating the propagation and summation in the Manchester Carry Chain (MCC), we avoided this problem and proposed an innovative quantum higher radix adder. Specifically, this circuit can be divided into two parts.
Firstly, in the higher radix part, we use Gidney's Logical-And\cite{Craig_Gidney_2017} and Selinger's Multi-control Toffoli construction method\cite{T_depth_one_2013} to implement a general quantum higher radix circuit.
Secondly, in the MCC part, we chose the Brent-Kung structure as the carry path and Gidney's RCA as the sum path after carefully analysing all possible carry propagation structures and sum paths.
By integrating the quantum higher radix and MCC parts, we propose a quantum higher radix adder.

This work improves the efficiency of quantum circuits at different scales by proposing an innovative higher radix adder circuit. It is hoped that our paper can contribute to a deeper understanding of the great potential of multi-control Toffoli gates and higher radix strategy in improving the performance of quantum arithmetic circuits.

The remainder of this paper is divided into eight parts. The next part (\secn \ref{s:adder-prev}) introduces prominent previous research works. Following that, the second part (\secn \ref{s:adder-stru}) describes the implementation details of the higher radix layer and the whole structure of the quantum higher radix adder. We present the evaluation results in the third part (\secn \ref{s:adder-resu}). We conclude the paper in thereafter (\secn \ref{s:adder-conc}), though some additional information/discussion/example can be found in the three subsequent parts (Appendices \ref{appendix6-adder}, \ref{appendix7-adder} and \ref{best-r}).

%% file: document/2-0-previous-works.tex
In this section, we describe the previous relevant research. These related works can be divided into the following three types.

Firstly, we start by introducing some important quantum adders based on the structure of classical adders.
Over the past few decades, a large number of quantum adder designs have been published. These works can be divided into quantum RCA and quantum CLA.
The structures of quantum RCA are very simple, but the computation time tends to grow linearly with the qubit number. In 1995, Vedral \cite{VBE} et al. designed a simple reversible quantum adder that is based on the classical RCA structure. However, VBE adder requires linear ancilla qubits, which incurs a very large cost.
In order to reduce the cost of the VBE adder, Cuccaro \cite{Cuccaro} et al. designed a new structure that only requires one ancilla qubit with lower T-depth and T-count.
Unlike the quantum RCA, the quantum CLA tends to be more efficient and is capable of performing addition in exponential time.
In 2004, Draper \cite{Draper08} et al. borrowed the classical CLA structure and then designed a quantum logarithmic T-depth adder based on the Brent-Kung structure \cite{Brent-Kung}.  
Takahashi \cite{Takahashi08, Takahashi09} et al. made further optimizations on the Draper's adder, resulting in the designs of a variety of cheaper quantum CLA adders.

Besides, there are also some quantum adders that use the unique properties of qubits so that they cannot be implemented in the classical world. 
For example, Gideney \cite{Craig_Gidney_2017} proposed a Logical-And structure based on the properties of qubits which significantly reduces the cost required to build a pair of Toffolis. Compared to Cuccaro's adder, Gideney's design greatly reduces the T-count and T-depth required to construct a quantum RCA adder.
In the later sections, Gideney‘s RCA will be used as our sum path in the MCC part.

Most importantly, we draw the higher radix strategy from classical adders.
In 2000, Gurkaynak et al. \cite{classic_higher_radix} found that higher radix adders that have larger fan-in and fan-out tend to be more efficient than radix-2 CLAs. 
However, the higher radix idea has not been applied in the quantum world so far. 
This is perhaps due to the fact that qubits are not copyable, which means the fan-out of the propagation part in a quantum CLA cannot be larger than 2. Therefore, we cannot construct higher radix quantum adders with higher fan-out in the quantum world.
To solve this problem, we borrow the idea of separating the carry chain and sum chain from the classical MCC adder. Specifically, our quantum adder is devided into two parts. The first part is called the carry path, which is used to calculate specific intermediate carry bits. We use Brent-Kung structure \cite{Brent-Kung} to construct the carry path. The second part is called the sum path, which is used to compute the result of the addition according to those specific intermediate carry bits obtained from the carry path. In this paper, we use multiple parallel carry-select adders (CSAs) or quantum RCAs to construct the sum path. 
Overall we show that various quantum adders can be categorized as qubit count(QC), T-count, T-depth, similar to Harris’ classification of classical adders \cite{Harris}.

%% file: document/3-0-description.tex
In this section, we first introduce C$_n$NOT which is an important basic component of the quantum higher radix adder.
Specifically, our method is divided into three steps.
In the first step, we describe how to construct the higher radix layer based on the C$_n$NOT gate.
In the second and third steps, the Brent-Kung structure and Gidney’s RCA are chosen as the Carry path and Sum path by analyzing five carry-propagate structures and two sum structures, respectively.
At the end of this section, we describe how to construct the overall higher radix circuit in detail.

%% file: document/3-1-basic-component.tex
C$_n$NOT is a basic component of the quantum higher radix adder.
In order to construct a cheap C$_n$NOT gate, we use and optimize Gidney's Logical-And structure. 
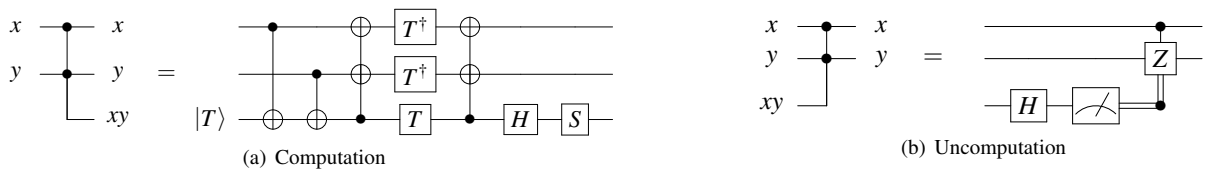
\begin{figure}[!ht]{
    \centering
    \subfigure[Computation\label{fig:Logical-And: Computation}]{ \resizebox{0.46\linewidth}{!}{
\ensuremath{  \Qcircuit @C=1em @R=.5em {
      x && \ctrl{2} &\qw &x&&  &&&&\ctrl{2}&\qw&\targ&\gate{T^\dag}&\targ&\qw&\qw&\qw \\
      y && \ctrl{1} &\qw &y&&= &&&&\qw&\ctrl{1}&\targ& \gate{T^\dag}&\targ&\qw&\qw&\qw \\
    &&&\qw & xy &&&&&\lstick{\ket{T}}&\targ&\targ&\ctrl{-2}& \gate{T}&\ctrl{-2} &  \gate{H}&  \gate{S}&\qw\\&
}}   }
}
    \hspace{1.6cm}
\smallskip
    \centering
    \subfigure[Uncomputation\label{fig:b}]{
    \resizebox{0.33\linewidth}{!}{
\ensuremath{ \Qcircuit @C=1em @R=.5em {
     x && \ctrl{2} &\qw &x&&  &&&\qw&\qw&\ctrl{1}&\qw  \\
     y && \ctrl{1} &\qw &y&&= &&&\qw&\qw&\gate{Z}&\qw \\
     xy&&\qw&& &&&&&\gate{H}& \meter & \control \cw\cwx\\&
}}}   }
    \caption{Gidney's Logical-And structure.\label{fig:Logical_And}}
}
\end{figure}
 \begin{equation}
\ket{T}= 
\begin{cases}
    \frac{1}{\sqrt{2}}(\ket{0}+e^{i\pi/4}\ket{1})& \text{~if~} ancilla=\ket 0; \\
    \frac{1}{\sqrt{2}}(\ket{0}-e^{i\pi/4}\ket{1}) & \text{~if~} ancilla=\ket 1 \label{e:radix}\\
\end{cases}
\end{equation}

In Gidney's paper \cite{Craig_Gidney_2017}, the first formula in Equation \eqref{e:radix} is used to define the special state \ket{T} in the Logical-And structure (Figure \ref{fig:Logical-And: Computation}). According to it, we can apply a Hadamard gate first and then a T gate on an ancilla with state \ket{0} to obtain \ket{T}. 
However, we found that it is also possible to use ancilla with state \ket{1} instead of \ket{0} to construct this structure. The related formula is shown in the lower part of Equation \eqref{e:radix}. In brief, we can perform operations such as NOT before Logical-And structure, which can be used to reduce qubits required for our quantum adder in later sections.
After expanding the scope of application of Logical-And, we then describe the specific structure and decomposition method of the proposed multi-control Toffoli gate.
\begin{table}[!ht]
\caption{\label{table_CnNOT}%
Cost of C$_n$NOT.
}
\centering
\begin{tabular}{|c|c|c|c|}
\hline
\textrm{Structure}&
\textrm{T-count}&
\textrm{T-depth}&
\textrm{QC}\\
\hline
C$_2$NOT& $TC$ & $TD$ & $Q$\\\hline
C$_3$NOT & $4+TC$ & $2+TD$ & $2+Q$\\\hline
C$_4$NOT& $8+TC$ & $3+TD$ & $4+Q$\\\hline
C$_5$NOT & $12+TC$ & $3+TD$ & $6+Q$\\\hline
C$_6$NOT& $16+TC$ & $4+TD$ & $8+Q$\\\hline
C$_{n+2}$NOT& $4n+TC$ & $2+\left \lfloor\log n\right \rfloor+TD$ & $2n+Q$\\\hline
\multicolumn{4}{c}{$TC = 7$, $TD = 3$, $Q = 3$.}
\end{tabular}

\end{table}
\begin{figure}[!ht]{
    \centering
\smallskip
    \centering
    \subfigure[C$_3$NOT: Unpaired\label{fig:c}]{
    \resizebox{.46\textwidth}{!}{
    \ensuremath{ \Qcircuit @C=0.85em @R=1.2em {
     &\ctrl{3} &\qw&&&&&&\lstick{\ket{x}}&\ctrl{4}&\qw&\ctrl{4}&\qw&&\lstick{\ket{x}}  &&&&&\lstick{\ket{x}}&\ctrl{4}&\qw&\ctrl{4}&\qw&&\lstick{\ket{x}} \\
     &\ctrl{2} &\qw&&&&&&\lstick{\ket{y}}&\ctrl{3}&\qw&\ctrl{3}&\qw&&\lstick{\ket{y}}  &&&&&\lstick{\ket{y}}&\ctrl{3}&\qw&\ctrl{3}&\qw&&\lstick{\ket{y}} \\
     &\ctrl{1} &\qw&&=&&&&\lstick{\ket{z}}&\qw&\ctrl{1}&\qw&\qw&&\lstick{\ket{z}}  &=&&&&\lstick{\ket{z}}&\qw&\ctrl{1}&\qw&\qw&&\lstick{\ket{z}} \\
     &\targ &\qw&&&&&&\lstick{\ket{0}}&\qw&\targ&\qw&\qw&&&\lstick{\ket{xyz}}  &&&&\lstick{\ket{0}}&\qw&\targ&\qw&\qw&&&\lstick{\ket{xyz}}\\
     &&&&&&&&\lstick{\ket{0}}&\targ&\ctrl{-1}&\targ&\qw&&\lstick{\ket{0}}&&&&&  &&\ctrl{-1}&\qw
     \gategroup{3}{11}{5}{11}{1em}{--}
     \\\\
}}}}
\smallskip
    \centering
    \subfigure[C$_3$NOT: Computation\label{fig:d}]{\resizebox{.46\textwidth}{!}{
    \ensuremath{ \Qcircuit @C=0.85em @R=1.2em {
     x&&\ctrl{3} &\qw&&&&&&\lstick{\ket{x}}&\ctrl{4}&\qw&\qw&&\lstick{\ket{x}}  &&&&&\lstick{\ket{x}}&\ctrl{4}&\qw&\qw&&\lstick{\ket{x}} \\
     y&&\ctrl{2} &\qw&&&&&&\lstick{\ket{y}}&\ctrl{3}&\qw&\qw&&\lstick{\ket{y}}  &&&&&\lstick{\ket{y}}&\ctrl{3}&\qw&\qw&&\lstick{\ket{y}} \\
     z&&\ctrl{1} &\qw&&=&&&&\lstick{\ket{z}}&\qw&\ctrl{1}&\qw&&\lstick{\ket{z}}  &=&&&&\lstick{\ket{z}}&\qw&\ctrl{1}&\qw&&\lstick{\ket{z}} \\
     && &\qw&xyz&&&&&\lstick{\ket{0}}&\qw&\targ&\qw&&&\lstick{\ket{xyz}}  &&&&\lstick{\ket{0}}&\qw&\targ&\qw&&&\lstick{\ket{xyz}}\\
     &&&&&&&&&\lstick{\ket{0}}&\targ&\ctrl{-1}&\qw&& &&&&&  &&\ctrl{-1}&\qw\\&
}}
    }}
\hfil
\smallskip
    \centering
    \subfigure[C$_3$NOT: Uncomputation\label{fig:e} ]{\resizebox{.46\textwidth}{!}{
    \ensuremath{ \Qcircuit @C=0.85em @R=1.2em {
    x&&\ctrl{3} &\qw&&&&&&\lstick{\ket{x}}&\qw&\ctrl{4}&\qw&&\lstick{\ket{x}}  &&&&&\lstick{\ket{x}}&\qw&\ctrl{4}&\qw&&\lstick{\ket{x}} \\
    y&&\ctrl{2} &\qw&&&&&&\lstick{\ket{y}}&\qw&\ctrl{3}&\qw&&\lstick{\ket{y}}  &&&&&\lstick{\ket{y}}&\qw&\ctrl{3}&\qw&&\lstick{\ket{y}} \\
    z&&\ctrl{1} &\qw&&=&&&&\lstick{\ket{z}}&\ctrl{1}&\qw&\qw&&\lstick{\ket{z}}  &=&&&&\lstick{\ket{z}}&\ctrl{1}&\qw&\qw&&\lstick{\ket{z}} \\
     xyz&& \qw&&&&&&&\lstick{\ket{xyz}}&\targ&\qw&\qw&&\lstick{\ket{0}}&&  &&&\lstick{\ket{xyz}}&\targ&\qw&\qw&&\lstick{\ket{0}}\\
     &&&&&&&&&\lstick{\ket{0}}&\ctrl{-1}&\targ&\qw&&\lstick{\ket{0}}&&&&&  &\ctrl{-1}&\qw\\&
}}}
}
    \caption{Construction of multi-control Toffoli.
        \label{fig: construct_multi_control_Toffoli}}

}
\end{figure}
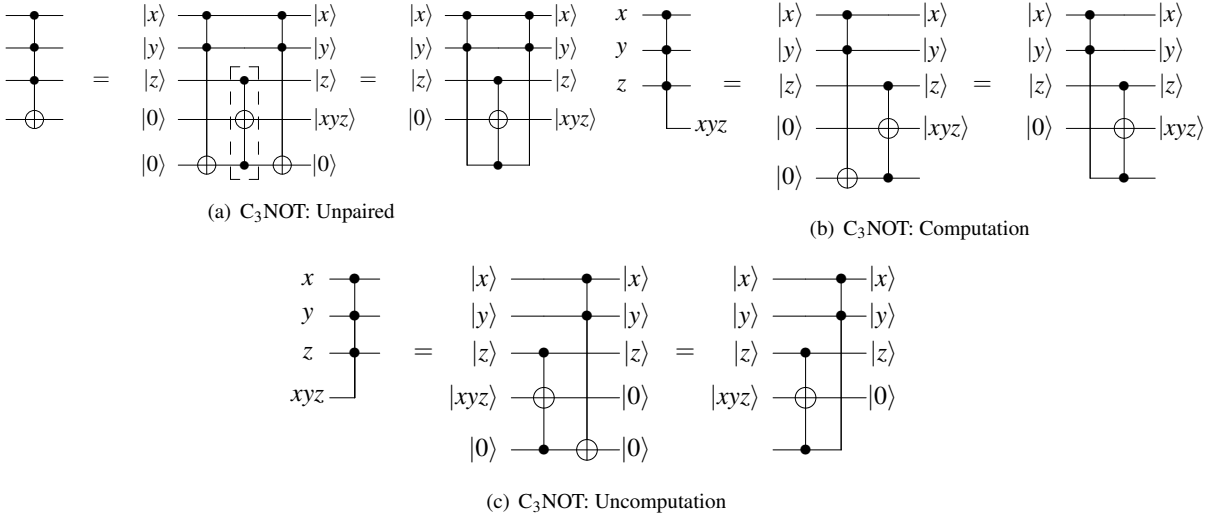

As shown in Table \ref{table_CnNOT}, we found that as the number of control qubits increases, the C$_n$NOT gate can effectively reduce the average T-count, T-depth, and QC per control qubit. In order to reduce the cost required by the circuit, we use multi-control Toffolis in this paper.
In 2013, Selinger \cite{T_depth_one_2013} proposed a general method for constructing  C$_n$NOT gates using Clifford gates and T gates. This paper optimizes Selinger's general method by referring to Logical-And structure \cite{Craig_Gidney_2017} and another Toffoli decomposition method \cite{CSWAP1}.
Figure \ref{fig: construct_multi_control_Toffoli} shows the specific structure of our multi-control Toffoli gate. 
Based on Logical-And structure, we  take C$_3$NOT as an example to show how to construct multi-control Toffoli gates.
In Figure \ref{fig:c}, we show how to construct an unpaired C$_3$NOT gate with Clifford + T gates. For a pair of C$_3$NOT gates, Figure \ref{fig:d} and \ref{fig:e} introduce the computation and uncomputation circuits, respectively. 
For Figures \ref{fig:c}, \ref{fig:d} and \ref{fig:e}, the first circuits from left to right are the original designs, and the second circuits show how to use Toffolis to decompose circuits in the first column. Besides, the third circuits use Logical-And structure to decompose the original circuits.

Specifically, using computation and uncomputation structures of Logical-And to decompose the rest of the Toffoli gates except the middle one can effectively reduce the T-count and T-depth of our multi-control Toffolis. As a result, the efficiency of the whole circuit is increased.

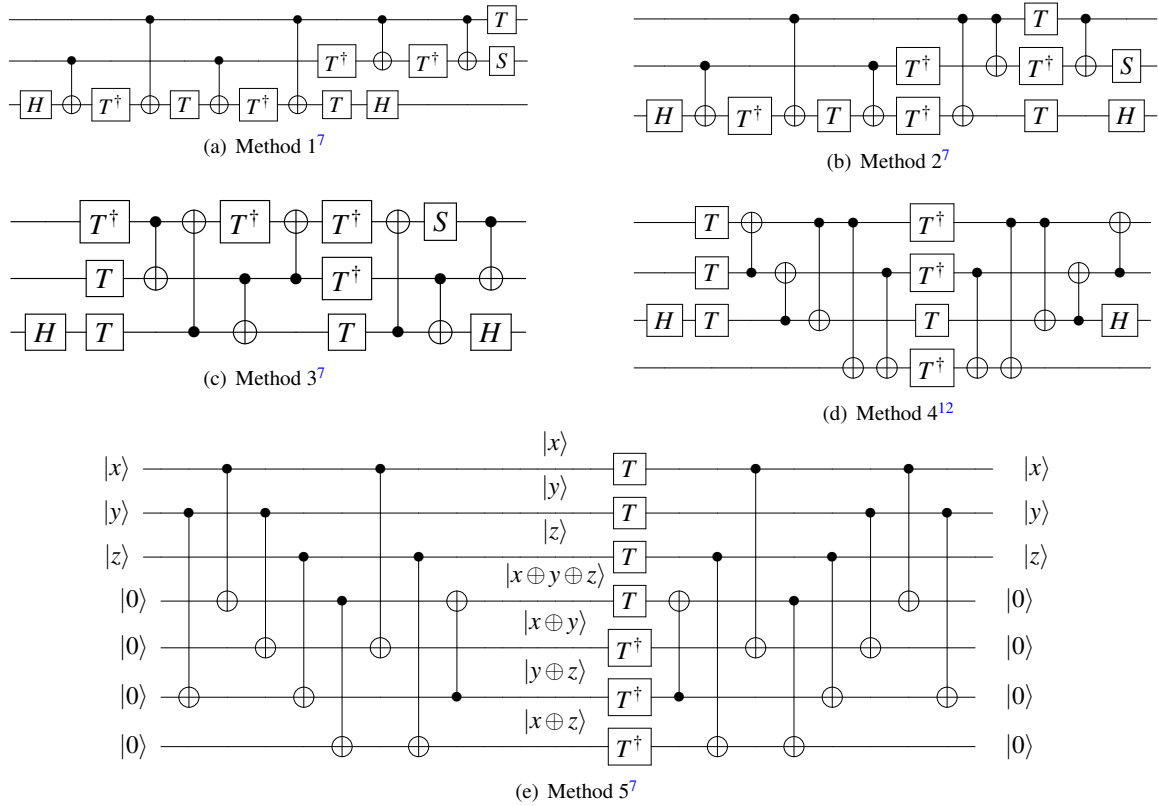
\begin{figure}[!ht]
    \centering
    \subfigure[Method 1 \cite{T_depth_one_2013}\label{fig: Decomposition_Toffoli-a}]{
    \resizebox{.4\textwidth}{!}{
    \ensuremath{ \Qcircuit @C=0.5em @R=.5em {
&\qw&\qw&\qw&\ctrl{2}&\qw&\qw&\qw&\ctrl{2}&\qw&\ctrl{1}&\qw&\ctrl{1}&\gate{T}&\qw\\
&\qw&\ctrl{1}&\qw&\qw&\qw&\ctrl{1}&\qw&\qw&\gate{T^\dag}&\targ&\gate{T^\dag}&\targ&\gate{S}&\qw\\ 
&\gate{H}&\targ&\gate{T^\dag}&\targ&\gate{T}&\targ&\gate{T^\dag}&\targ&\gate{T}&\gate{H}&\qw&\qw&\qw&\qw\\&&
} } }}
    \hfil
    \subfigure[Method 2\cite{T_depth_one_2013}\label{fig: Decomposition_Toffoli-b}]{
    \resizebox{.4\textwidth}{!}{
    \ensuremath{ \Qcircuit @C=0.5em @R=.5em {
&\qw &\qw &\qw&\ctrl{2}&\qw&\qw&\qw&\ctrl{2}&\ctrl{1}&\gate{T}&\ctrl{1}&\qw&\qw\\
&\qw &\ctrl{1} &\qw&\qw&\qw&\ctrl{1}&\gate{T^\dag}&\qw&\targ&\gate{T^\dag}&\targ&\gate{S}&\qw \\
&\gate{H}&\targ&\gate{T^\dag}&\targ&\gate{T}&\targ&\gate{T^\dag}&\targ&\qw&\gate{T}&\qw&\gate{H}&\qw\\&&
}}}}
    \subfigure[Method 3 \cite{T_depth_one_2013}\label{fig: Method 3}]{
    \resizebox{.4\textwidth}{!}{
    \ensuremath{ \Qcircuit @C=0.5em @R=.5em {
&\qw&\gate{T^\dag}&\ctrl{1}&\targ&\gate{T^\dag}&\targ&\gate{T^\dag}&\targ&\gate{S}&\ctrl{1}&\qw \\
&\qw&\gate{T}&\targ&\qw&\ctrl{1}&\ctrl{-1}&\gate{T^\dag}&\qw&\ctrl{1}&\targ&\qw \\
&\gate{H}&\gate{T}&\qw&\ctrl{-2}&\targ&\qw&\gate{T}&\ctrl{-2}&\targ&\gate{H}&\qw\\&&}} }}
    \hfil
    \subfigure[Method 4\cite{CSWAP1}\label{fig: Decomposition_Toffoli-d}]{
    \resizebox{.4\textwidth}{!}{
    \ensuremath{ \Qcircuit @C=0.5em @R=.5em {
&\qw&\gate{T}&\targ&\qw&\ctrl{2}&\ctrl{3}&\qw&\gate{T^\dag}&\qw&\ctrl{3}&\ctrl{2}&\qw&\targ&\qw\\
&\qw&\gate{T}&\ctrl{-1}&\targ&\qw&\qw&\ctrl{2}&\gate{T^\dag}&\ctrl{2}&\qw&\qw&\targ&\ctrl{-1}&\qw\\
&\gate{H}&\gate{T}&\qw&\ctrl{-1}&\targ&\qw&\qw&\gate{T}&\qw&\qw&\targ&\ctrl{-1}&\gate{H}&\qw\\
&\qw&\qw&\qw&\qw&\qw&\targ&\targ&\gate{T^\dag}&\targ&\targ&\qw&\qw&\qw&\qw
\\&&}} }}
    \subfigure[Method 5 \cite{T_depth_one_2013}\label{fig: Decomposition_Toffoli-e}]
    {\resizebox{0.66\linewidth}{!}
    {
\ensuremath{ \Qcircuit @C=0.7em @R=.5em {
\lstick{\ket{x}}&\qw&\qw&\ctrl{3}&\qw&\qw&\qw&\ctrl{4}&\qw&\qw&\qw&\qw&\qw& \qw&\ustick{\ket{x}}\qw&\qw&\qw&\gate{T}&\qw&\qw&\ctrl{4}&\qw&\qw&\qw&\ctrl{3}&\qw&\qw&\qw&\rstick{\ket{x}}\\
\lstick{\ket{y}}&\qw&\ctrl{4}&\qw&\ctrl{3}&\qw&\qw&\qw&\qw &\qw&\qw&\qw&\qw&\qw&\ustick{\ket{y}}\qw&\qw&\qw&\gate{T}&\qw&\qw&\qw&\qw&\qw&\ctrl{3}&\qw&\ctrl{4}&\qw&\qw&\rstick{\ket{y}}\\
\lstick{\ket{z}}&\qw&\qw&\qw&\qw&\ctrl{3}&\qw&\qw&\ctrl{4}&\qw&\qw&\qw&\qw&\qw&\ustick{\ket{z}}\qw&\qw&\qw&\gate{T}&\qw&\ctrl{4}&\qw&\qw&\ctrl{3}&\qw&\qw&\qw&\qw&\qw&\rstick{\ket{z}}\\
&\lstick{\ket{0}}&\qw&\targ&\qw&\qw&\ctrl{3}&\qw&\qw&\targ&\qw&\qw&\qw&\qw&\ustick{\ket{x\oplus y\oplus z}}\qw&\qw&\qw&\gate{T}&\targ&\qw&\qw&\ctrl{3}&\qw&\qw&\targ&\qw&\qw&\rstick{\ket{0}}\\
&\lstick{\ket{0}}&\qw&\qw&\targ&\qw&\qw&\targ&\qw&\qw&\qw&\qw&\qw&\qw&\ustick{\ket{x\oplus y}}\qw&\qw&\qw&\gate{T^\dag}&\qw&\qw&\targ&\qw&\qw&\targ&\qw&\qw&\qw&\rstick{\ket{0}}\\
&\lstick{\ket{0}}&\targ&\qw&\qw&\targ&\qw&\qw&\qw&\ctrl{-2}&\qw&\qw&\qw&\qw&\ustick{\ket{y\oplus z}}\qw&\qw&\qw&\gate{T^\dag}&\ctrl{-2}&\qw&\qw&\qw&\targ&\qw&\qw&\targ&\qw&\rstick{\ket{0}}\\
&\lstick{\ket{0}}&\qw&\qw&\qw&\qw&\targ&\qw&\targ&\qw&\qw&\qw&\qw&\qw&\ustick{\ket{x\oplus z}}\qw&\qw&\qw&\gate{T^\dag}&\qw&\targ&\qw&\targ&\qw&\qw&\qw&\qw&\qw&\rstick{\ket{0}}\\&&
}}}
}
    \caption{Different methods to decompose unpaired Toffoli with Clifford $+$ T gates.\cite{T_depth_one_2013,CSWAP1}
    }
    \label{fig: Decomposition_Toffoli}
\end{figure}

As shown in Figure \ref{fig: Decomposition_Toffoli}, there are five general decomposition methods for the middle unpaired Toffoli gate. \cite{CSWAP1,T_depth_one_2013}According to Table \ref{table_Toffoli_decomposition}, we can find that all decomposition methods use 7 T gates. Since the proposed higher radix adder has a high QC, we do not wish to introduce more ancilla qubits by using decomposition methods with ancilla bits. Although the T-depth of the quantum circuit that we obtain using Method 5 to decompose one unpaired Toffoli is only 1, it will introduce 4 additional ancillae, which will greatly increase qubits to construct a CLA, so we do not choose this decomposition method. Similarly, Method 4 also introduces extra ancilla qubits. Among Methods 1, 2, and 3, Method 3 has the highest efficiency because it has the smallest T-depth. Hence, we choose Method 3 to decompose all the unpaired Toffolis in our work.
\begin{table}[!ht]
\caption{\label{table_Toffoli_decomposition}%
Summary table of Toffoli decomposition.
}
\centering
\begin{tabular}{|c|c|c|c|}
\hline
\textrm{Decomposition}&
\textrm{T-count}&
\textrm{T-depth}&
\textrm{Ancilla Count}\\
\hline
Method 1 (Figure \ref{fig: Decomposition_Toffoli-a}) & 7 & \textcolor[rgb]{1,0,0}{6} & \textcolor[rgb]{0,0,1}{0}\\\hline
Method 2 (Figure \ref{fig: Decomposition_Toffoli-b})& 7 & 4 & \textcolor[rgb]{0,0,1}{0}\\\hline
Method 3 (Figure \ref{fig: Method 3})& 7 & 3 & \textcolor[rgb]{0,0,1}{0}\\\hline
Method 4 (Figure \ref{fig: Decomposition_Toffoli-d})& 7 & 2 & 1\\\hline
Method 5 (Figure \ref{fig: Decomposition_Toffoli-e})& 7 &\textcolor[rgb]{0,0,1}{1} & \textcolor[rgb]{1,0,0}{4}\\\hline
\end{tabular}
\end{table}

%% file: document/3-2-higher-radix-layer.tex
In this section, we introduce the higher  radix strategy and implementation details of how to apply it to quantum circuits.
Radix is the bit-width of each CLA block in carry look-ahead adders. For convenience, we use $r$ to represent radix. In the classical world, the higher radix adder has shown its advantages \cite{classic_higher_radix}. By increasing the radix, an adder with higher fan-in and fan-out is constructed, which can reduce the propagation time of propagates ($p$) as well as generates ($g$), and improve the efficiency of classical adder.

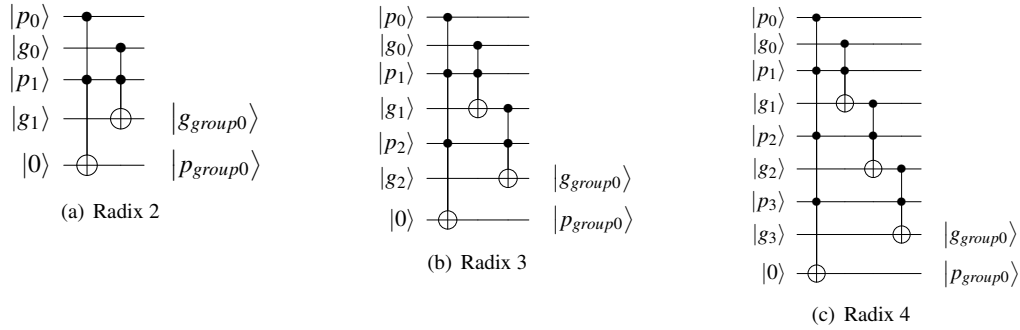
\begin{figure}[!ht]
    \centering
    \subfigure[Radix 2\label{fig:radix_2}]{
    \resizebox{.08\textwidth}{!}{
    \ensuremath{ \Qcircuit @C=0.5em @R=1.em {
\lstick{\ket{p_0}}&\ctrl{4}&\qw&\qw\\
\lstick{\ket{g_0}}&\qw&\ctrl{2}&\qw\\
\lstick{\ket{p_1}}&\ctrl{2}&\ctrl{1}&\qw\\
\lstick{\ket{g_1}}&\qw&\targ&\qw&\rstick{\ket{g_{group0}}}\\
\lstick{\ket{0}}&\targ&\qw&\qw&\rstick{\ket{p_{group0}}}\\ &
}}   }
    }
     %
    ~\hfil~
    \subfigure[Radix 3\label{fig:radix_3}]{
    \resizebox{.09\textwidth}{!}{
\ensuremath{ \Qcircuit @C=0.5em @R=1.em {
\lstick{\ket{p_0}}&\ctrl{6}&\qw&\qw&\qw\\
\lstick{\ket{g_0}}&\qw&\ctrl{2}&\qw&\qw\\
\lstick{\ket{p_1}}&\ctrl{4}&\ctrl{1}&\qw&\qw\\
\lstick{\ket{g_1}}&\qw&\targ&\ctrl{2}&\qw\\
\lstick{\ket{p_2}}&\ctrl{2}&\qw&\ctrl{1}&\qw\\
\lstick{\ket{g_2}}&\qw&\qw&\targ&\qw&\rstick{\ket{g_{group0}}}\\
\lstick{\ket{0}}&\targ&\qw&\qw&\qw&\rstick{\ket{p_{group0}}}\\&
}}}}
    ~\hfil~
    \subfigure[Radix 4\label{fig:radix_4}]{
    \resizebox{.11\textwidth}{!}{
\ensuremath{ \Qcircuit @C=0.5em @R=1.em {
\lstick{\ket{p_0}}&\ctrl{8}&\qw&\qw&\qw&\qw\\
\lstick{\ket{g_0}}&\qw&\ctrl{2}&\qw&\qw&\qw\\
\lstick{\ket{p_1}}&\ctrl{6}&\ctrl{1}&\qw&\qw&\qw\\
\lstick{\ket{g_1}}&\qw&\targ&\ctrl{2}&\qw&\qw\\
\lstick{\ket{p_2}}&\ctrl{4}&\qw&\ctrl{1}&\qw&\qw\\
\lstick{\ket{g_2}}&\qw&\qw&\targ&\ctrl{2}&\qw\\
\lstick{\ket{p_3}}&\ctrl{2}&\qw&\qw&\ctrl{1}&\qw\\
\lstick{\ket{g_3}}&\qw&\qw&\qw&\targ&\qw&\rstick{\ket{g_{group0}}}\\
\lstick{\ket{0}}&\targ&\qw&\qw&\qw&\qw&\rstick{\ket{p_{group0}}}\\&
}}
}}
    \caption{Quantum circuits for higher radix layer. \label{fig: higher_radix_structure}}
\end{figure}

We firstly take the radix 2 CLA as an example to calculate the sum of two binary numbers $a$ and $b$. We need to get the carry $c$ after 2-step calculations.
In the first step, we calculate the $p$ and $g$ of each bit. For the $i$\textsuperscript{th} bit, we use the formulas $1$ and $2$ to compute the corresponding $p_i$ and $g_i$, respectively.
In the second step, after computing the propagation of $p$ and $g$ to obtain the intermediate quantities $P$ and $G$, we can calculate the carry $c$. 
Here, we assume that $j$ is less than or equal to $i$. 
Equations \eqref{e:radix-3}, \eqref{e:radix-4} and \eqref{e:radix-5} describe how to propagate $p$ and $g$ from bit $i$ to bit $j$. 
After using the formulas \eqref{e:radix-6}, we are able to obtain the final carry $c$.
The standard notations of 
$\oplus$ for logical XOR, $\cdot$, $+$, $\circ$ for logical AND,  logical OR and propagation, respectively are here.
\begin{eqnarray}
&p_i&= a_i \oplus b_i\label{e:radix-1}\\
&g_i&=a_i \cdot b_i\label{e:radix-2}\\
&(G_{0:0}, P_{0:0})&=(g_0, p_0)\label{e:radix-3}\\
&(G_{0:i}, P_{0:i})&=(g_i, p_i)\circ (G_{0:i-1}, P_{0:i-1})\label{e:radix-4}\\
&(g_x, p_x) \circ (g_y, p_y)&=(g_x+p_x\cdot g_y, p_x \cdot p_y)\label{e:radix-5}\\
&c_i&=g_i+p_i \cdot c_{i-1}\label{e:radix-6}
\end{eqnarray}
As shown in Figure \ref{fig: higher_radix_structure}, we use multi-control Toffoli gates to construct the quantum propagation structures with radix 2, 3, and 4 respectively. Similarly, the higher radix strategy also reduces the number of propagation layers of $p$ and $g$, thereby reducing T-count required for the addition operations, and further effectively reducing the operation time.
However, due to the extremely high cost of fan-out larger than 2, the quantum higher radix strategy proposed only focuses on increasing the fan-in of the adder.


In this paper, we try to construct a quantum higher radix adder, which means we need to construct the higher radix layer which is based on multi-control Toffolis for the propagation of $p$ and $g$. 
%
Figures \ref{fig:radix_3} and \ref{fig:radix_4} show the specific circuits for the higher radix structure with radix 3 and radix 4 as respective examples.

%% file: document/3-3-carry-path.tex
\begin{figure*}[!ht]
    \centering
    \subfigure[J. Sklansky and Ladner-Fisher\label{fig: 1-Sklansky}\label{fig:3-LF}]{
    \includegraphics[width=.475\linewidth]{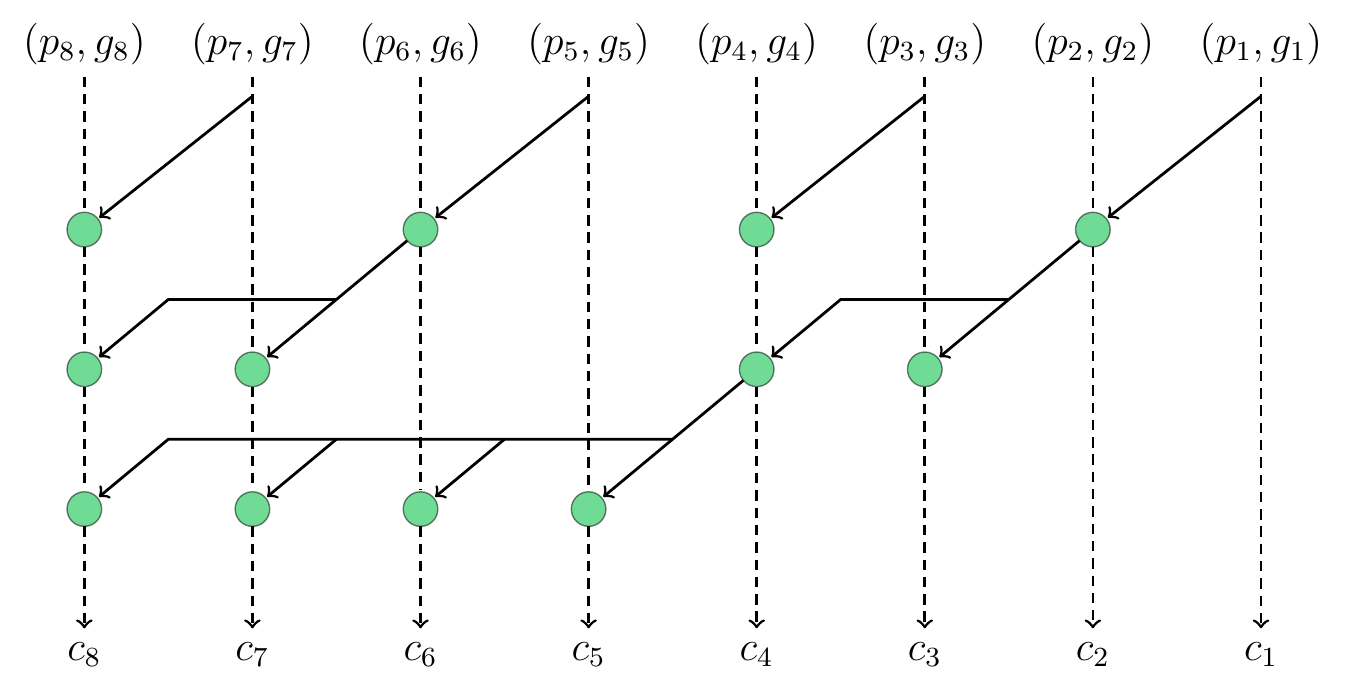}
    }
    \hfill
    \subfigure[Kogge-Stone\label{fig:2-KS}]{
    \includegraphics[width=.475\linewidth]{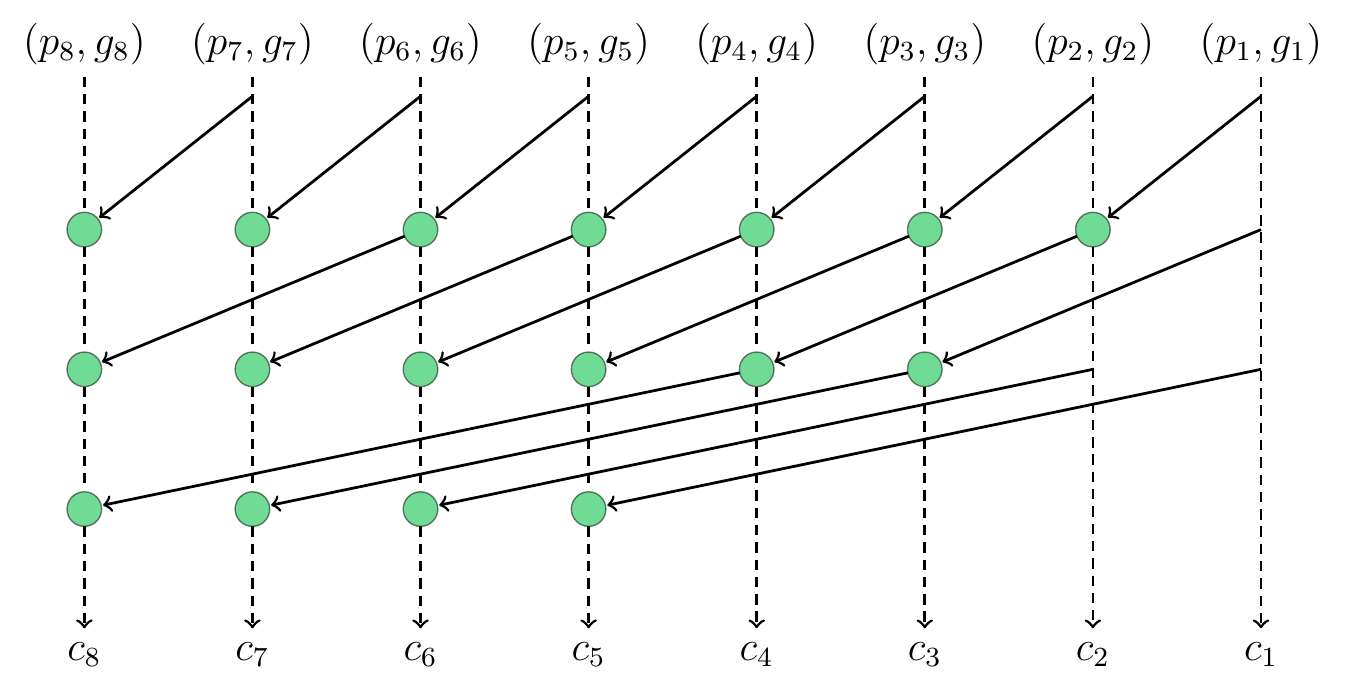}
    }
    \centering
    \hspace{10cm}
    \subfigure[Brent-Kung\label{fig:4-BK}]{
    \includegraphics[width=.475\linewidth]{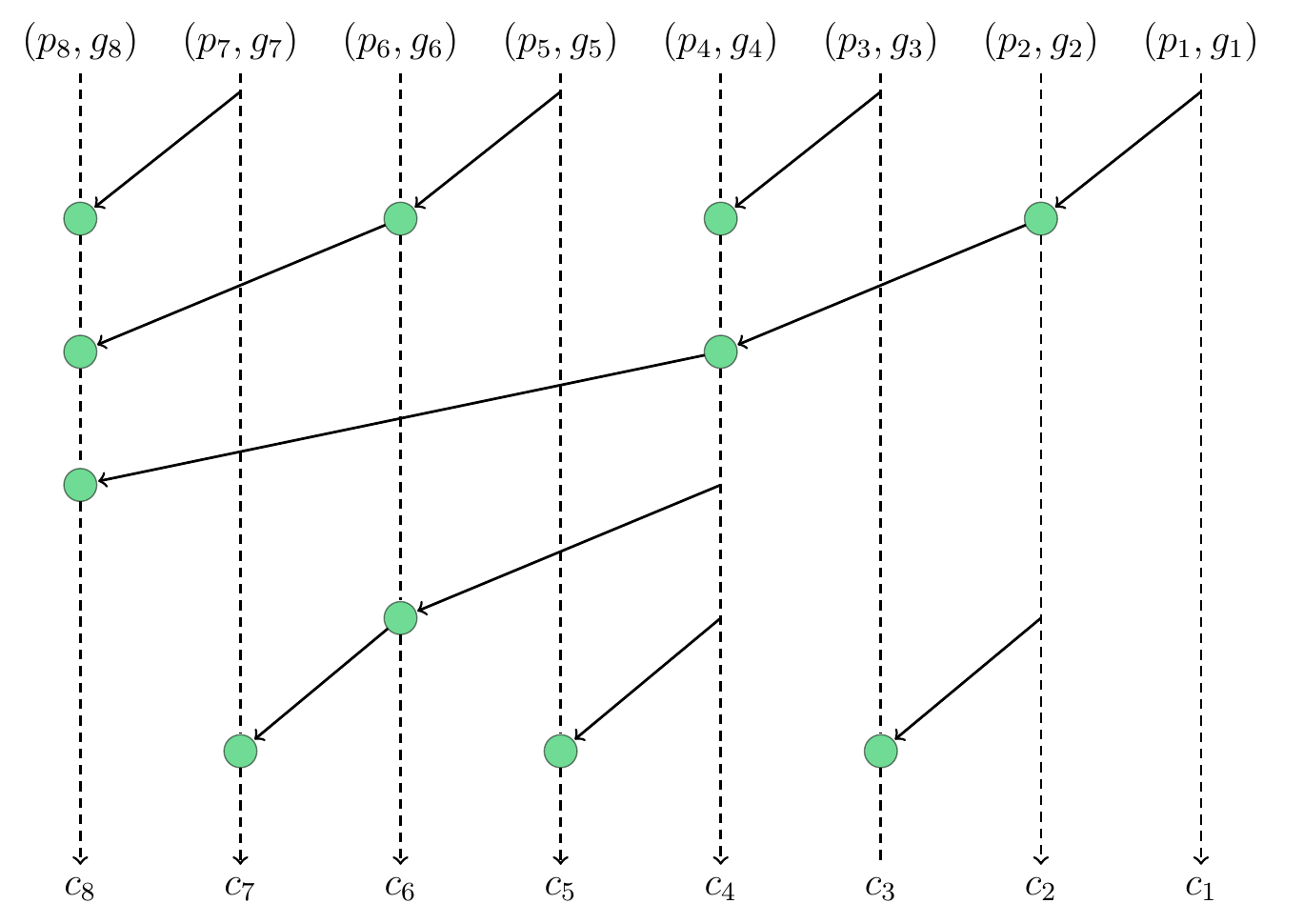}
    }
    \hfill
    \subfigure[Han-Carlson\label{fig:5-HC}]{\includegraphics[width=0.475\linewidth, ]{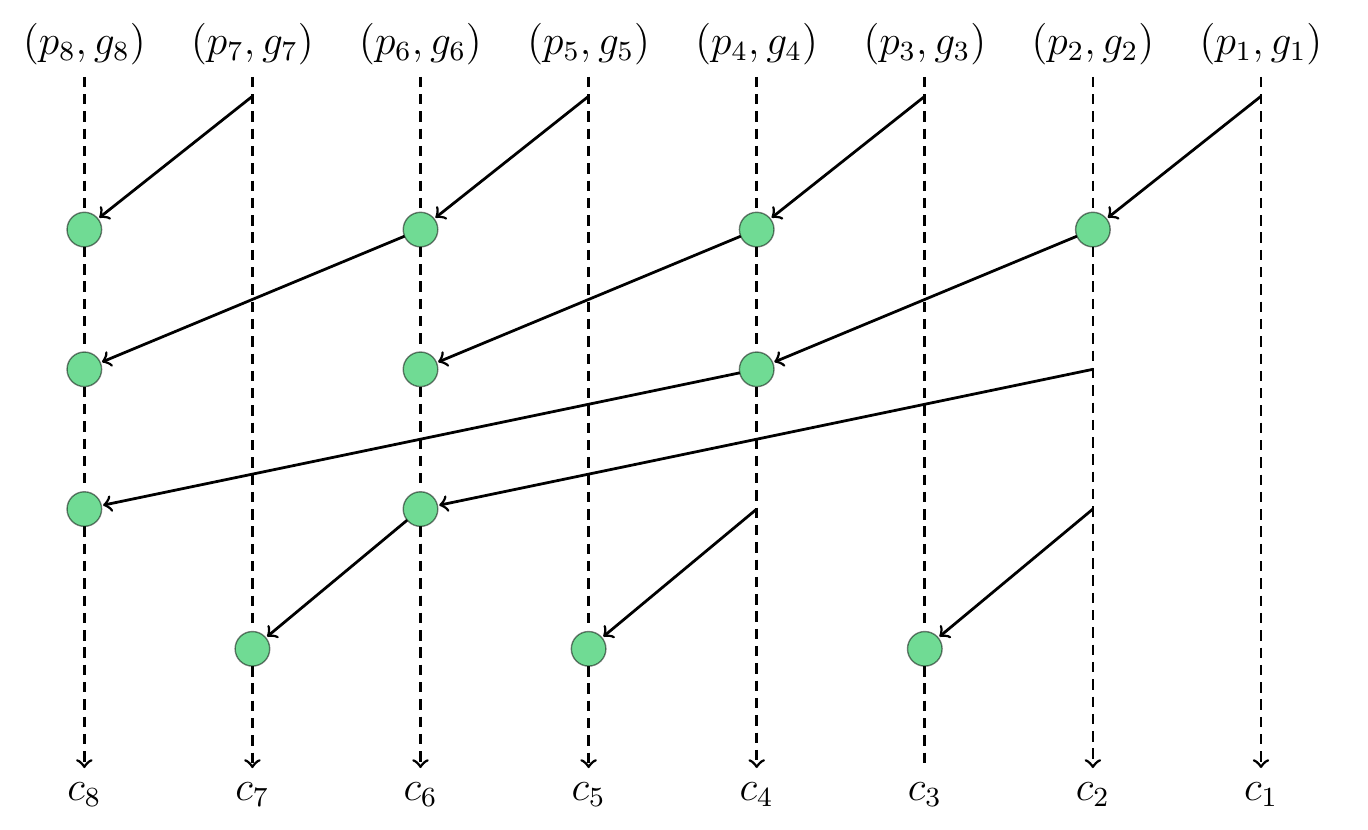}}
    
    \caption{Carry-propagate structures, where the green node represents one propagate operation.
    \label{fig:CLA}}
\end{figure*}

In this section, we describe the details of our carry path. 
Through the calculation in the previous section, we have already obtained $p$ and $g$ for each bit. In order to get the carries, we need to select a particular carry-propagate structure as the carry path to propagate $p$ and $g$. %
As shown in Figure \ref{fig:CLA}, five different propagation structures are proposed in the literature, which are subsequently discussed. 
\begin{figure*}[ht]
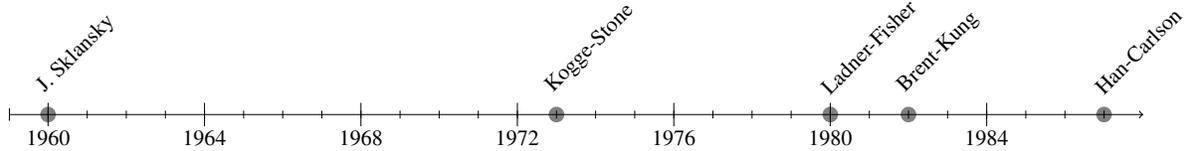

    \centering
       \resizebox{.9\textwidth}{!}{
\begin{chronology}[4]{1959}{1987}{\linewidth}
\event{1960}{\textcolor{black}{J. Sklansky}}
\event{1973}{\textcolor{black}{Kogge-Stone}}
\event{1980}{\textcolor{black}{Ladner-Fisher}}
\event{1982}{\textcolor{black}{Brent-Kung}}
\event{1987}{\textcolor{black}{Han-Carlson}}
\end{chronology}}
    \caption{Chronology of publication of carry-propagate structures. \label{fig:timeline2}}    
\end{figure*}

\begin{itemize}

    \item \textbf{Sklansky.} In 1960, J. Sklansky \cite{Sklansky} proposed a conditional CLA adder with high fan-out nodes and minimal depth. The structure of it is shown in Figure \ref{fig: 1-Sklansky}.
    \item \textbf{Kogge-Stone.} The Kogge-Stone structure \cite{Kogge-Stone} was published in 1973, which has a low depth but a high number of nodes. The structure is shown in Figure \ref{fig:2-KS}.
    \item \textbf{Ladner-Fisher.}  As shown in Figure \ref{fig:3-LF}, the topology of Ladner-Fisher \cite{Ladner-Fischer} looks the same as the Sklanskly structure. Hence, it also has low depth but high fan-out nodes. However, there are some differences between these two structures in the application.
    \item \textbf{Brent-Kung.} The Brent-Kung structure \cite{Brent-Kung} is one of the most important propagation structures. Compared to other structures, this structure has a very small number of nodes as well as low fan-in and fan-out, despite having a large logic depth. Therefore, it is widely used in quantum CLA designs.
    \item \textbf{Han-Carlson.} In 1987, the Han-Carlson structure was first proposed \cite{Han-Carlson}. In order to improve the overall efficiency of the propagation, it combines the Brent-Kung and Kogge-Stone structures together.

\end{itemize}

The propagation operations are the main cost of the carry path. Furthermore, since qubit can not be copied, the carry-propagate structures with fan-out larger than 2 introduce additional cost.
Therefore, as shown in Figure \ref{fig:CLA}, the Brent-Kung structure which has the smallest number of propagate operations and low fan-out is selected as the carry path for the $p$ and $g$ propagation in our paper.

%% file: document/3-4-sum-path.tex
We then discuss the implementation details of the sum path. As described in the previous sections, the final sum can be calculated by feeding carries into the sum path.
For the sum path, we can choose between CSA and RCA. 

\begin{itemize}
    \item \textbf{RCA.} The general structure of RCA is shown in Figure \ref{fig: Sumpath_1-RCA}. As discussed in \secn \ref{s:adder-prev}, there are various quantum RCAs have been proposed so far. In this paper, we choose the most efficient Gidney adder which has the minimum T-count and T-depth as our RCA structure. 
    
    \item \textbf{CSA.} As shown in Figure \ref{fig: Sumpath_2-CSA}, the Carry Select adder consists of two Ripple Carry adders and one select circuit. 
    
    In the first part, we built two quantum RCAs with the same structure. The input carry bit of them are set to $0$ and $1$. Therefore, using these sub-circuits, we can obtain the sum when the inputs are $0$ and $1$, respectively. In both the classical and quantum worlds, this part can be computed in parallel with the carry path, thus effectively reducing the time cost.
    
    In the second part, we construct a select sub-circuit.  Suppose we know that the real input carry is $c$, which can only be $0$ or $1$. Depending on $c$, the sum calculated by the corresponding RCA is then chosen as the final result. More specifically, when $c$ is equal to $0$, we choose the sum of the quantum RCA whose input carry is $0$ as the final sum. Similarly, when $c=1$, the final result is the sum of quantum RCA whose input carry is $1$.
    
    It is worth noting that the quantum CSA contains expensive CSWAP gates which can be decomposed by Clifford+T gates. As shown in Figure \ref{fig: CSWAP}, there are 2 decomposition methods. For Method 1 \cite{CSWAP1}, the CSWAP gate is decomposed into a Clifford+T circuit with 7 T gates and T-depth of 4. If we use Method 2 (this is adopted from Quipper documentation (\url{https://www.mathstat.dal.ca/~selinger/quipper/doc/Quipper-Libraries-GateDecompositions.html}.)), the CSWAP gate is decomposed into a circuit with 7 T gates incurring the T-depth of 3. 
    
    In this paper, we use CSA1 to denote the CSA structure whose CSWAPs are decomposed by Method 1. Similarly, CSA2 represents the CSA whose CSWAPs are decomposed by Method 2.

\end{itemize}

After determining the design details of these sum paths, we performed a systematic analysis of their performance. As shown in Table \ref{tab:sum_path}, the RCA structure is always cheaper than the CSA structures in terms of T-count, T-depth and QC. Therefore, we choose Gidney's RCA as our sum path.
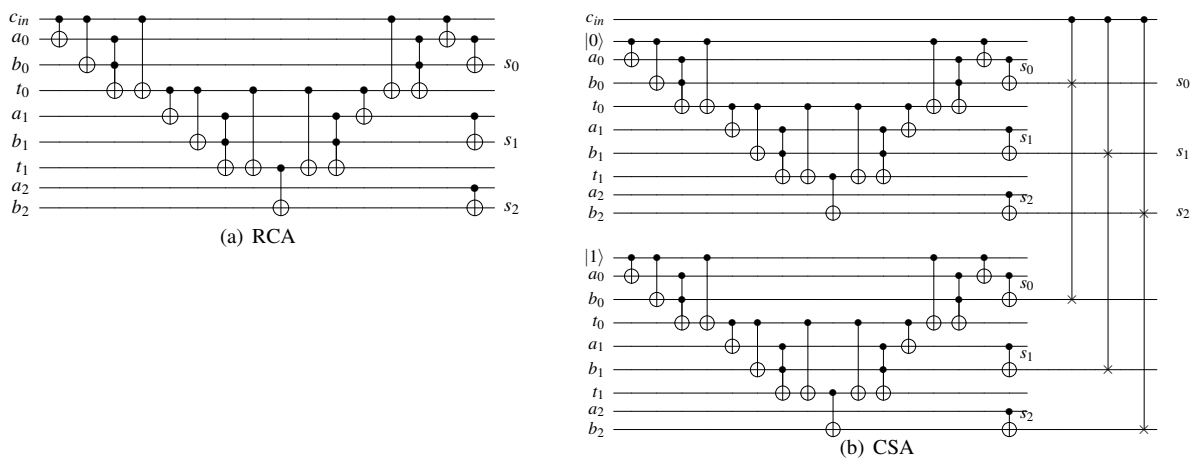
\begin{figure}[!ht]
    \centering
    \subfigure[RCA\label{fig: Sumpath_1-RCA}]{
    \resizebox{.36\textwidth}{!}{
    \ensuremath{ \Qcircuit @C=.6em @R=.5em {
&\lstick{c_{in}}&\ctrl{1}&\ctrl{2}&\qw&\ctrl{3}&\qw&\qw&\qw&\qw&\qw&\qw&\qw&\qw&\ctrl{3}&\qw&\ctrl{1}&\qw&\qw\\
&\lstick{a_0}&\targ&\qw&\ctrl{2}&\qw &\qw&\qw&\qw&\qw&\qw&\qw&\qw&\qw &\qw&\ctrl{2}&\targ&\ctrl{1}&\qw\\
&\lstick{b_0}&\qw&\targ&\ctrl{1}&\qw &\qw&\qw&\qw&\qw&\qw&\qw&\qw&\qw &\qw&\ctrl{1}&\qw&\targ&\rstick{s_0}\qw\\
&\lstick{t_0}&\qw&\qw&\targ&\targ&\ctrl{1}&\ctrl{2}&\qw&\ctrl{3}&\qw&\ctrl{3}&\qw&\ctrl{1}&\targ&\targ&\qw&\qw&\qw\\
&\lstick{a_1}&\qw&\qw&\qw&\qw&\targ&\qw&\ctrl{2}&\qw&\qw&\qw&\ctrl{2}&\targ&\qw&\qw&\qw&\ctrl{1}&\qw\\
&\lstick{b_1}&\qw&\qw&\qw&\qw&\qw&\targ&\ctrl{1}&\qw&\qw&\qw&\ctrl{1}&\qw&\qw&\qw&\qw&\targ&\rstick{s_1}\qw\\
&\lstick{t_1}&\qw&\qw&\qw&\qw&\qw&\qw&\targ&\targ&\ctrl{2}&\targ&\targ&\qw&\qw&\qw&\qw&\qw&\qw\\
&\lstick{a_2}&\qw&\qw&\qw&\qw&\qw&\qw&\qw&\qw&\qw&\qw&\qw&\qw&\qw&\qw&\qw&\ctrl{1}&\qw\\
&\lstick{b_2}&\qw&\qw&\qw&\qw&\qw&\qw&\qw&\qw&\targ&\qw&\qw&\qw&\qw&\qw&\qw&\targ&\rstick{s_2}\qw
\\&
}}
    }}
    \hfil
    \subfigure[CSA \label{fig: Sumpath_2-CSA}]{
    \resizebox{.43\textwidth}{!}{
    \ensuremath{ \Qcircuit @C=.6em @R=.5em {
&\lstick{c_{in}}&\qw&\qw&\qw&\qw&\qw&\qw&\qw&\qw&\qw&\qw&\qw&\qw&\qw&\qw&\qw&\qw&\qw  &\qw&\qw&\qw  &\ctrl{16}&\qw&\qw&\ctrl{19}&\qw&\qw&\ctrl{22}&\qw \\
&\\
&\lstick{\ket{0}}&\ctrl{1}&\ctrl{2}&\qw&\ctrl{3}&\qw&\qw&\qw&\qw&\qw&\qw&\qw&\qw&\ctrl{3}&\qw&\ctrl{1}&\qw&\qw  &&&\\
&\lstick{a_0}&\targ&\qw&\ctrl{2}&\qw &\qw&\qw&\qw&\qw&\qw&\qw&\qw&\qw &\qw&\ctrl{2}&\targ&\ctrl{1}&\qw  &&&\\
&\lstick{b_0}&\qw&\targ&\ctrl{1}&\qw &\qw&\qw&\qw&\qw&\qw&\qw&\qw&\qw &\qw&\ctrl{1}&\qw&\targ&\ustick{s_0}\qw  &\qw&\qw&\qw&\qswap&\qw&\qw&\qw&\qw&\qw&\qw&\qw &\rstick{s_0}\\
&\lstick{t_0}&\qw&\qw&\targ&\targ&\ctrl{1}&\ctrl{2}&\qw&\ctrl{3}&\qw&\ctrl{3}&\qw&\ctrl{1}&\targ&\targ&\qw&\qw&\qw  &&&\\
&\lstick{a_1}&\qw&\qw&\qw&\qw&\targ&\qw&\ctrl{2}&\qw&\qw&\qw&\ctrl{2}&\targ&\qw&\qw&\qw&\ctrl{1}&\qw  &&&\\
&\lstick{b_1}&\qw&\qw&\qw&\qw&\qw&\targ&\ctrl{1}&\qw&\qw&\qw&\ctrl{1}&\qw&\qw&\qw&\qw&\targ&\ustick{s_1}\qw   &\qw&\qw&\qw &\qw&\qw&\qw&\qswap&\qw&\qw&\qw&\qw&\rstick{s_1}\\
&\lstick{t_1}&\qw&\qw&\qw&\qw&\qw&\qw&\targ&\targ&\ctrl{2}&\targ&\targ&\qw&\qw&\qw&\qw&\qw&\qw  &&&\\
&\lstick{a_2}&\qw&\qw&\qw&\qw&\qw&\qw&\qw&\qw&\qw&\qw&\qw&\qw&\qw&\qw&\qw&\ctrl{1}&\qw  &&&\\
&\lstick{b_2}&\qw&\qw&\qw&\qw&\qw&\qw&\qw&\qw&\targ&\qw&\qw&\qw&\qw&\qw&\qw&\targ&\ustick{s_2}\qw  &\qw&\qw&\qw &\qw&\qw&\qw&\qw&\qw&\qw&\qswap&\qw&\rstick{s_2}
\\&&&&&&&&&&&&&&&&&&&&&\\&\\&\\
&\lstick{\ket{1}}&\ctrl{1}&\ctrl{2}&\qw&\ctrl{3}&\qw&\qw&\qw&\qw&\qw&\qw&\qw&\qw&\ctrl{3}&\qw&\ctrl{1}&\qw&\qw  &&&\\
&\lstick{a_0}&\targ&\qw&\ctrl{2}&\qw &\qw&\qw&\qw&\qw&\qw&\qw&\qw&\qw &\qw&\ctrl{2}&\targ&\ctrl{1}&\qw  &&&\\
&\lstick{b_0}&\qw&\targ&\ctrl{1}&\qw &\qw&\qw&\qw&\qw&\qw&\qw&\qw&\qw &\qw&\ctrl{1}&\qw&\targ&\ustick{s_0}\qw  &\qw&\qw&\qw &\qswap&\qw&\qw&\qw&\qw&\qw&\qw&\qw\\
&\lstick{t_0}&\qw&\qw&\targ&\targ&\ctrl{1}&\ctrl{2}&\qw&\ctrl{3}&\qw&\ctrl{3}&\qw&\ctrl{1}&\targ&\targ&\qw&\qw&\qw  &&&\\
&\lstick{a_1}&\qw&\qw&\qw&\qw&\targ&\qw&\ctrl{2}&\qw&\qw&\qw&\ctrl{2}&\targ&\qw&\qw&\qw&\ctrl{1}&\qw  &&&\\
&\lstick{b_1}&\qw&\qw&\qw&\qw&\qw&\targ&\ctrl{1}&\qw&\qw&\qw&\ctrl{1}&\qw&\qw&\qw&\qw&\targ&\ustick{s_1}\qw  &\qw&\qw&\qw&\qw&\qw&\qw&\qswap&\qw&\qw&\qw&\qw\\
&\lstick{t_1}&\qw&\qw&\qw&\qw&\qw&\qw&\targ&\targ&\ctrl{2}&\targ&\targ&\qw&\qw&\qw&\qw&\qw&\qw  &&&\\
&\lstick{a_2}&\qw&\qw&\qw&\qw&\qw&\qw&\qw&\qw&\qw&\qw&\qw&\qw&\qw&\qw&\qw&\ctrl{1}&\qw  &&&\\
&\lstick{b_2}&\qw&\qw&\qw&\qw&\qw&\qw&\qw&\qw&\targ&\qw&\qw&\qw&\qw&\qw&\qw&\targ&\ustick{s_2}\qw  &\qw&\qw&\qw  &\qw&\qw&\qw&\qw&\qw&\qw&\qswap&\qw
\\
}}}}
    \caption{Sum paths.\label{fig: Sum_path}}

\end{figure}
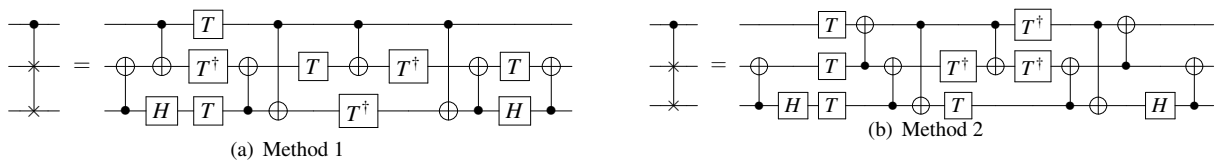
\begin{figure}[!ht]
    \centering
    \subfigure[Method 1\label{fig: Decomposition 1}]{
    \resizebox{.43\linewidth}{!}{
    \ensuremath{ \Qcircuit @C=.5em @R=.5em {
&\qw&\ctrl{1}&\qw&\qw&&&& &\qw&\ctrl{1}&\gate{T}&\qw&\ctrl{2}&\qw&\ctrl{1}&\qw&\ctrl{2}&\qw&\qw&\qw&\qw
\\
&\qw&\qswap&\qw&\qw&&=&& &\targ&\targ&\gate{T^\dag}&\targ&\qw&\gate{T}&\targ&\gate{T^\dag}&\qw&\targ&\gate{T}&\targ&\qw
\\
&\qw&\qswap\qwx&\qw&\qw&&&& &\ctrl{-1}&\gate{H}&\gate{T}&\ctrl{-1}&\targ&\qw&\gate{T^\dag}&\qw&\targ&\ctrl{-1}&\gate{H}&\ctrl{-1}&\qw\\
\\
}}}
}
\hfil
\subfigure[Method 2\label{fig: Decomposition 2}]{
       \resizebox{.43\linewidth}{!}{
    \ensuremath{ \Qcircuit @C=.5em @R=.5em {
&\qw&\ctrl{1}&\qw&\qw&&&& &\qw&\qw&\gate{T}&\targ&\qw&\ctrl{2}&\qw&\ctrl{1}&\gate{T^\dag}&\qw&\ctrl{2}&\targ&\qw&\qw&\qw\\
&\qw&\qswap&\qw&\qw&&=&& &\targ&\qw&\gate{T}&\ctrl{-1}&\targ&\qw&\gate{T^\dag}&\targ&\gate{T^\dag}&\targ&\qw&\ctrl{-1}&\qw&\targ&\qw\\
&\qw&\qswap\qwx&\qw&\qw&&&& &\ctrl{-1}&\gate{H}&\gate{T}&\qw&\ctrl{-1}&\targ&\gate{T}&\qw&\qw&\ctrl{-1}&\targ&\qw&\gate{H}&\ctrl{-1}&\qw\\
}}}}
    \caption{Decomposition methods for quantum CSWAP gate.\label{fig: CSWAP}}
\end{figure}


\begin{table}[!ht]
\caption{\label{tab:sum_path}%
Different structures of sum path ($r=n$). \\When calculating T-depth, we assume that Part 1 of the CSA has been completed when calculating the carry path. Therefore, for CSA1 and CSA2, T-depth equals to the T-depth of Part 2, which is the minimum T-depth of sum path for CSAs.}
\centering
\begin{tabular}{|c|c|c|c|}\hline
\textrm{Structure}&
\textrm{T-count}&
\textrm{T-depth}&
\textrm{QC}\\\hline
\hline
CSA1 & $11n-4$ & $4n$ & $6n+1$\\\hline
CSA2 & $11n-4$ & $3n$ & $6n+1$\\\hline
RCA  & \textcolor[rgb]{0,0,1}{$4n-4$} & \textcolor[rgb]{0,0,1}{$n$} & \textcolor[rgb]{0,0,1}{$3n$}\\
\hline
\end{tabular}
\end{table}


%% file: document/3-5-Overall.tex
\subsubsection*{One Higher Radix Layer}
\begin{figure}[!ht]
    \centering
    \subfigure[Classical version.\label{fig:r4BK1}]{\includegraphics[width=1.0\linewidth]{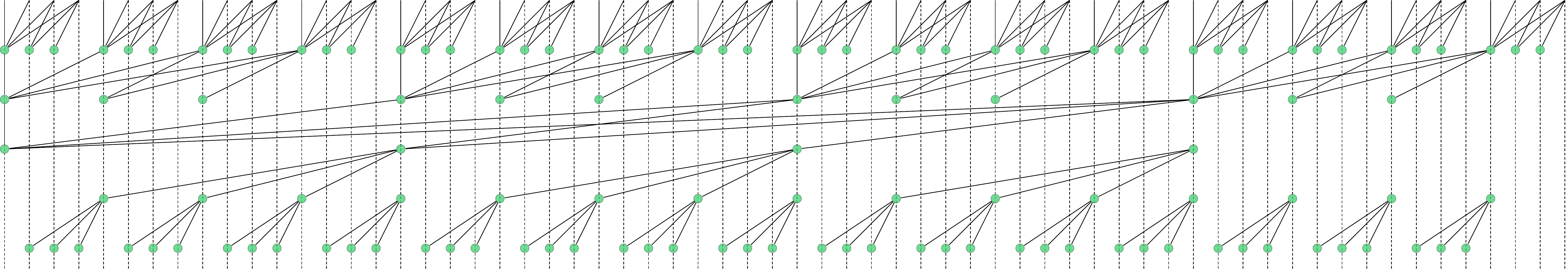}}
    \hspace{1.6cm}
    \subfigure[Quantum version. 
    \label{fig:r4BK2}]{\includegraphics[width=1.0\linewidth]{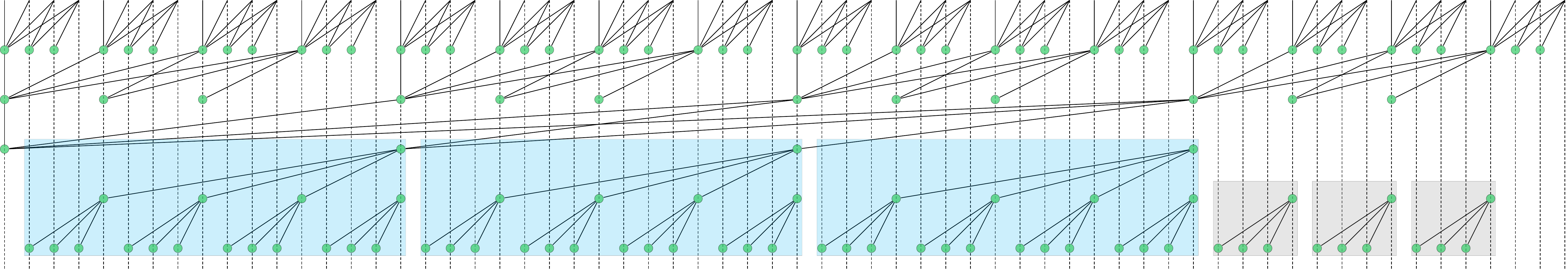}}
    
    \caption{Brent-Kung structure with radix 4. (The small bounding box represents radix-4 RCA, and the large bounding box represents radix-16 RCA.)\label{fig: r4BK}}
\end{figure}

In this paper, we only use one higher radix layer.
As shown in Figure \ref{fig:r4BK1} (This diagram was inspired from \url{https://web.stanford.edu/class/archive/ee/ee371/ee371.1066/lectures/lect_04.pdf}), the classical higher radix adder applies the higher radix strategy to every layer of the Brent-Kung tree, thus reducing the depth of computation from $log2^n$ to $log_r^n$ .

\begin{figure}[!ht]
    \centering
\includegraphics[width=.5\linewidth]{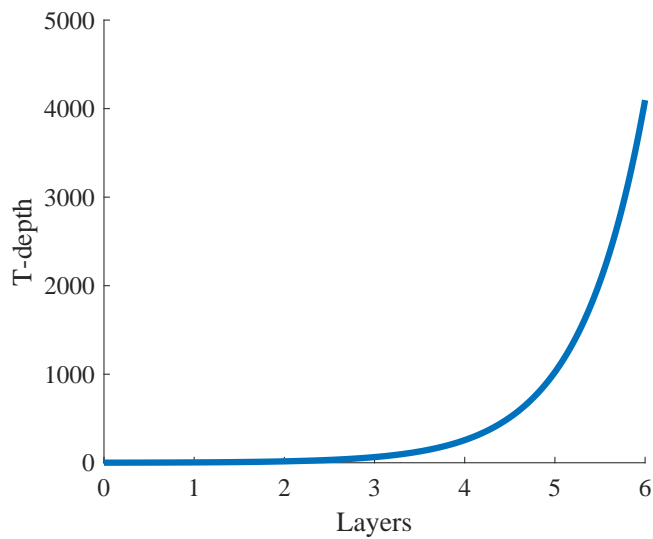}

    \caption{The T-depth of the sum path with increasing layers for the radix-4 higher radix adder. The formula represented by this curve is $O(T-depth)=r^{layers}$.\label{fig: layers}}
\end{figure}

In this part, we explain why do we not follow the same example from the classical computing. 
According to Figure \ref{fig: layers}, this is due to the fact that if the strategy is used at every layer, RCAs with large T-depth will be introduced in the sum path, which deprives our higher radix adder of the significant advantage of low T-depth.
Therefore, we recommend using the higher radix strategy only for the first few layers of the quantum Brent-Kung Tree.

\subsubsection*{Complete circuit} \label{ss:The whole circuit}
\begin{figure*}[!ht]
    
    \centering
    \includegraphics [height=0.915\textheight]{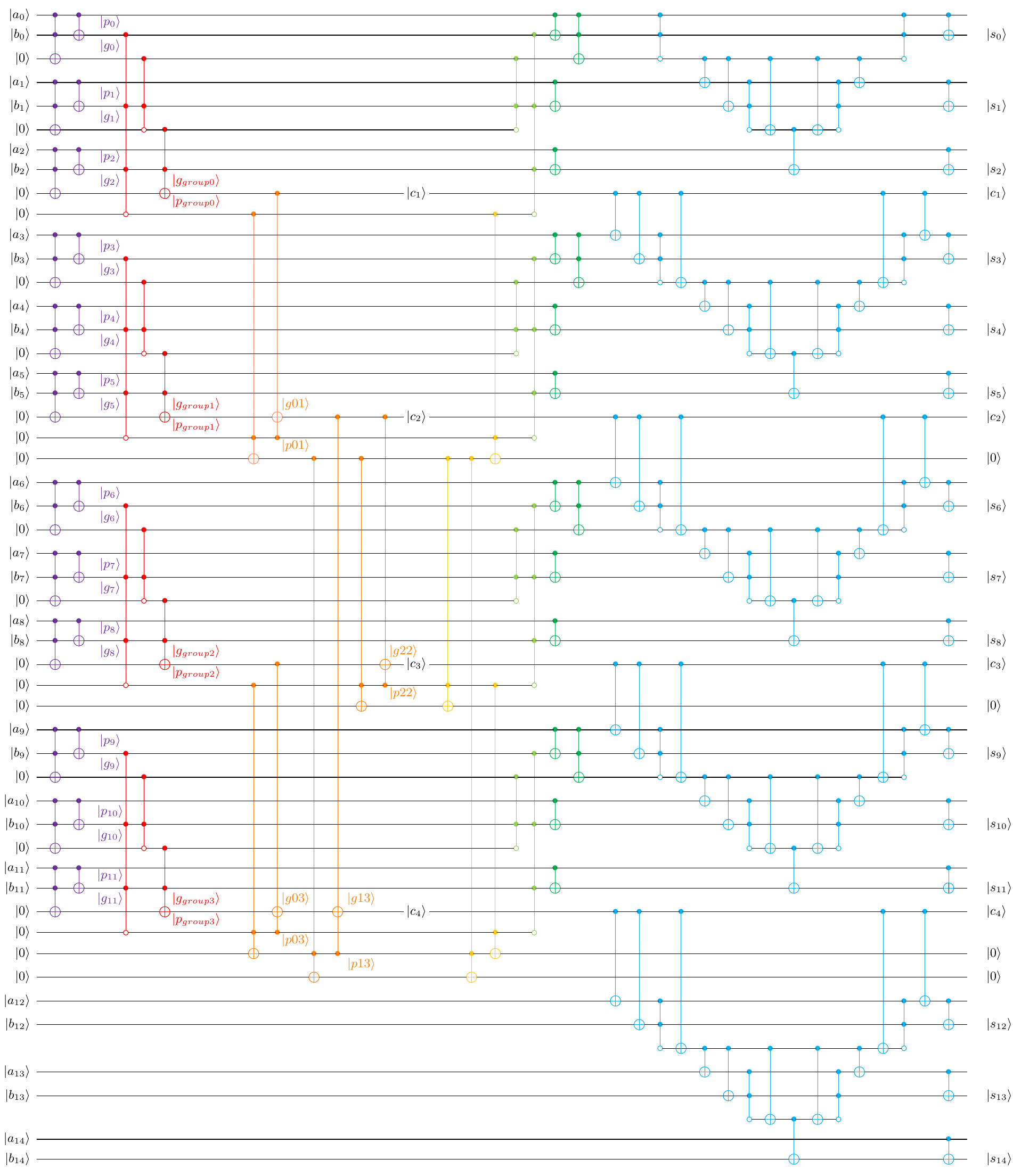}
    
    \caption{Radix-3 addition circuit with two 15-bit addends $a$ and $b$.\label{bits=15radix=3}
}

\end{figure*}

The structure of the proposed quantum higher radix circuit is shown in Figure \ref{bits=15radix=3}.
It can be divided into 7 stages.
\begin{itemize}
    \item \textbf{Notations.}
    The binary bit-width of the addends is denoted as $n$, and $r$ denotes the value of the radix. The binary expansion of the number $a$ is denoted as $a=a_{n-1}a_{n-2}\cdots a_0$, where $a_{n-1}$ is the most significant bit and $a_0$ is the least significant bit. 
    For circuit decomposition, paired Toffoli gates are decomposed into Logical-And, while unpaired Toffoli gates are decomposed using Method 3 shown in Figure \ref{fig: Decomposition_Toffoli}. Thus, $TC_3$ is equal to $7$ and $TD_3$ is equal to $3$.

    \item \textbf{Step 1.} In this step, our task is to calculate the $p$ and $g$.
    Since we do not need the most significant carry, no operation is performed on the most significant group.
    By taking $a_i$, $b_i$ as control qubits and an ancilla with initial state $\ket{0}$ as the controlled qubit, we apply the CCNOT gate to compute $g_i$ and then store it in the corresponding ancilla. 
    After that, we use $a_i$ as the control and $b_i$ as the controlled qubit to apply the CNOT gate. As a result, the corresponding $p_i$ is stored in the corresponding $b_i$ position. 
    Some Toffoli gates are unpaired in the whole circuit, we decompose those Toffoli gates by using Method 3.

    For convenience, $\alpha$ is introduced to denote the addend qubits in the most significant group.
    This step requires $TC_3\cdot(n-\alpha)$ T-count, $TD_3$ T-depth, and extra ancilla qubits are $3\cdot n-\alpha$.
 \begin{equation}
\alpha = 
\begin{cases}
        r & \text{~if~} n \pmod r = 0;\\
        n \pmod{r} &\text{~otherwise}
\end{cases}
\end{equation}

    \item \textbf{Step 2.}\label{step2-qadder} In the second step, we group the initially obtained $p$ and $g$ by using the higher radix structure.
Specifically, we construct the corresponding higher radix structure according to the method shown in Figure \ref{fig: higher_radix_structure}, and then apply it to the corresponding $g_i$ and $p_i$ calculated in step 1 to obtain $g_{group}$ and $p_{group}$.
Since the controlled qubits of the last Toffoli will be used to store the carry later, no uncomputation is performed on it. Hence, the last Toffoli is always unpaired, we only decompose it using Method 3, but decompose the rest of the Toffolis into Logical-And.

    For convenience, $\beta$ and $\rho$ are introduced to represent a complex intermediate variable for constructing multi-control Toffolis and the number of groups divided, respectively.
    In step 2, the required T-count is $\rho\cdot [TC_3+4\cdot(2\cdot r-3)]$, the T-depth is $ TD_3+r+\beta-1$, and extra ancilla is $\rho\cdot (r-1)$.
    \begin{equation}
\beta=
\begin{cases}
        0 & \text{~if~} r\leq 2;\\
        2+\lfloor\log (r-2)\rfloor &\text{~otherwise}
\end{cases}
\end{equation}
    \begin{equation}
\rho=
\begin{cases}
        \frac{n}{r}-1 & \text{~if~} n\pmod r=0;\\
        \lfloor\frac{n}{r} \rfloor &\text{~otherwise}
\end{cases}
\end{equation}

    \item \textbf{Step 3.} In step 3, we construct the Brent-Kung tree using the $p_{group}$ and $g_{group}$ processed by the higher radix structure to calculate the carry path. We tried using Logical-And here, but found no benefit. Hence, here all the Toffolis are decomposed by Method 3.

    In this step, the required T-count is $2\cdot TC_3 \cdot [2\cdot \rho -1 -\omega (\rho)-\lfloor\log (\rho)\rfloor]$, the T-depth is $TD_3\cdot( \left \lfloor\log (\rho)\right \rfloor+ \left \lfloor\log \frac{\rho}3\right \rfloor+2)$, and the number of extra ancilla qubits is $2\cdot \rho -1 - \omega(\rho)-\lfloor\log (\rho)\rfloor$.

    \item \textbf{Step 4.} In this step we uncompute the operation of calculating the intermediate $p$ in step 3. We repeat the calculation of all Toffolis for the intermediate variable $p$ in the reverse order.

    In step 4, the required T-count is  $ TC_3 \cdot [2\cdot \rho -1 -\omega (\rho)-\lfloor\log (\rho)\rfloor]$, the T-depth is $TD_3\cdot( \left \lfloor\log \rho\right \rfloor+ \left \lfloor\log \frac{\rho}3\right \rfloor+1)$. Here we do not need any extra ancilla qubit. 
    
    \item \textbf{Step 5.} Here we uncompute Step 2. We just repeat the same Toffolis from Step 2 in reverse order, except for the last one.
    
    Since we decompose them into Logical-And structures, no additional cost is needed in this step.

    
   \item \textbf{Step 6.} In step 6, we restore the original binary addends $a$ and $b$ by applying the NOT gate and Toffoli on $p$ and $g$. 
    In order to store the corresponding $b_i$ in the corresponding qubits, we apply the CNOT gate by taking $a_i$ as the control qubit and $b_i$ as the controlled qubit.
    Since the Logical-And that used previously introduces the measurement operation, so we then uncompute only the Toffoli gates which are applied to the least significant qubits of each group.
    
    In this step, the required T-count is $\rho\cdot TC_3$, and the T-depth is $TD_3$. We do not need any extra ancilla qubits. 
    
    \item \textbf{Step 7.} In this step, we construct the Gidney's RCA \cite{Craig_Gidney_2017} for every group to calculate the sum. 
    
    In step 7, the required T-count is $4\cdot (n-\lceil\frac{n}{r} \rceil)$, the T-depth is $r$, and the number of extra ancilla qubits is $\alpha-1+(r-2)\cdot \rho$.

\end{itemize}

An example of an addition operation performed by the higher radix adder is shown in Figure \ref{bits=15radix=3}. 
    We use seven colors to divide the whole circuit from step 1 to step 7 from left to right.  The radix of this adder is set to 3 and the inputs (i.e., addends) are two 15-bit binary numbers denoted by $a$ and $b$. By using this quantum circuit we can correctly get the sum of these two numbers.      In order to show the overall structure more clearly, we use $\circ$ to represent the controlled-NOT operation of the Logical-And structure in Figure \ref{bits=15radix=3}.

Interestingly, there are two special cases. When $r \geq n$, our adder is Gidney's RCA. When radix is equal to one, our adder is a simple CLA.
In summary, the overall cost of our circuit is shown below.
\begin{eqnarray}
    &\text{T-count}&=(8r+40)\cdot\left \lceil\frac{n}{r}\right \rceil+11n -72 -7\cdot(n-1)\pmod r -8r
    -21\omega(\left \lceil\frac{n}{r}\right \rceil-1)  -21\left \lfloor\log (\left \lceil\frac{n}{r}\right \rceil-1) \right \rfloor)\label{formula-TC}\\
&\text{T-depth}&= 
6\cdot (\left \lfloor\log (\left \lceil\frac{n}{r}\right \rceil-1) \right \rfloor +4+\left \lfloor\log (\frac{1}{3}(\left \lceil\frac{n}{r}\right \rceil-1)) \right \rfloor)
    +\left \lfloor\log(r-2) \right \rfloor +2r-5\label{formula-TD}\\
&\text{Qubit Count}&=
3n-1-2r-\omega(\left \lceil\frac{n}{r}\right \rceil-1) 
+(2r-1)\cdot\left \lceil\frac{n}{r}\right \rceil 
  -\left \lfloor\log (\left \lceil\frac{n}{r}\right \rceil-1) \right \rfloor\label{formula-QC}
\end{eqnarray}

It can be observed that the circuit structure of the quantum higher radix adder varies with radix. In the next section, we will discuss how radix affects the performance of our adder and compare it with other well-known work.

%% file: document/4-1-Experiment-1.tex
\begin{figure}[!ht]
    \centering
    \includegraphics[scale=0.66]{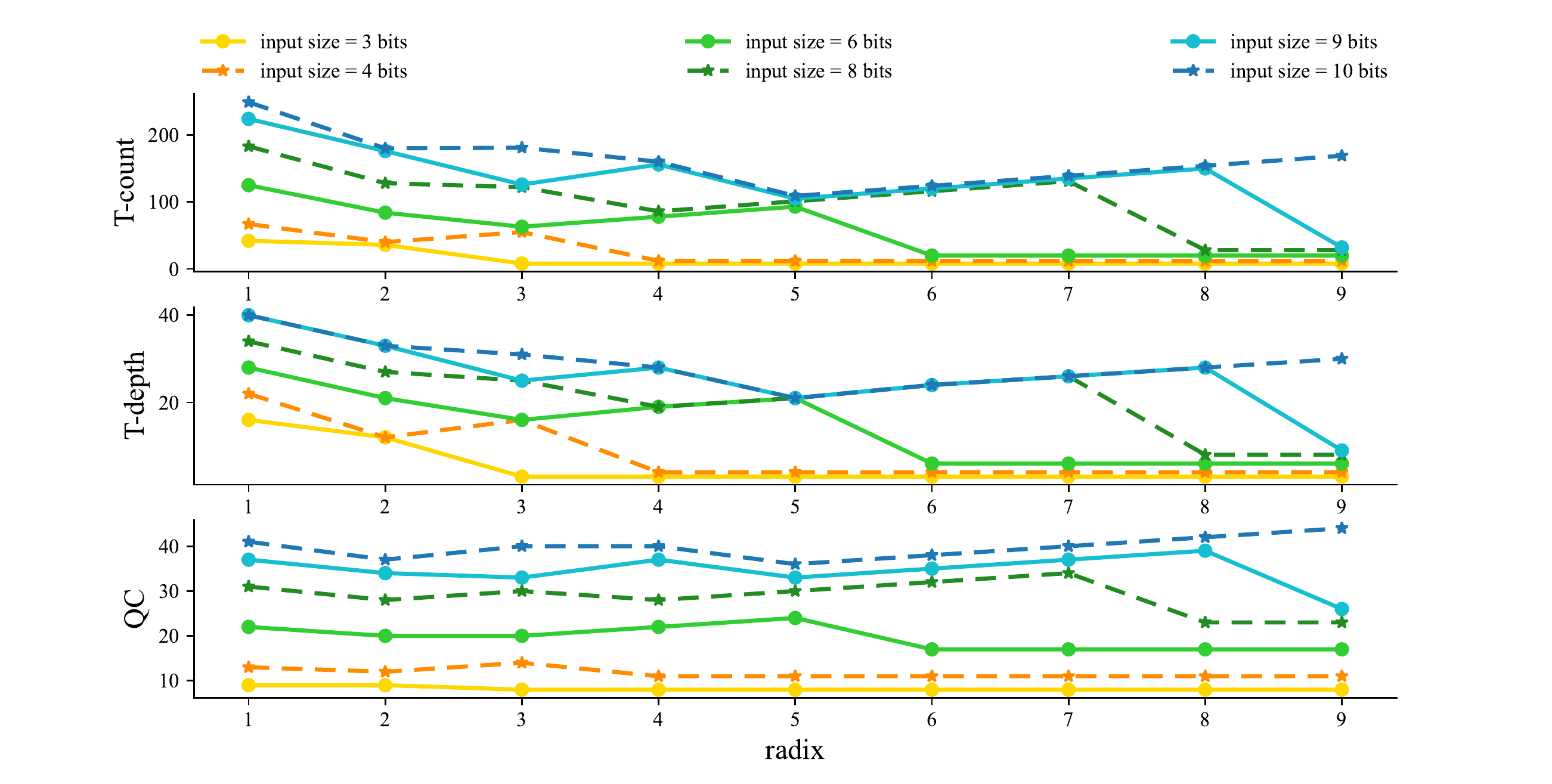}

    \caption{Comparision of the cost required by quantum higher radix adders with different radix and different sizes.\label{fig:fig_Compare_radix}}
\end{figure}

Figure \ref{fig:fig_Compare_radix} shows T-count, T-depth, and QC for nine different higher radix adders with radix from 1 to 9, respectively.
It is clear that when the radix is fixed, the performance of our adder varies for different input sizes. As the input size increases, the overall cost is higher, which means that the larger the input size, the more complex and expensive any adder tends to be. Since an increase in input size means an increase in the number of operations, this can directly result in an increase in circuit scale. For larger circuit, more expensive cost is often required in terms of T-depth, T-count and QC.

Interestingly, for a fixed input size, increasing the radix does not reduce the cost monotonically.
In this paper, the higher radix adder with $r$ equal to 1 is a CLA, while $r$ equal to 2 represents the higher radix adder without multi-control Toffolis. 
Compared to higher radix adder with $r$ equal to 1, the cost of it with radix 2 is reduced in T-count, T-depth and QC, which means that the higher radix layer can effectively optimize the circuit even without introducing multi-control Toffoli.
When $r$ is larger than 2, our adder is a hybrid of quantum RCA and CLA. T-count and T-depth decrease first and then increase as the radix increases in the range where $r$ is less than the input size. Meanwhile, QC increases steadily in the fluctuation. 
When $r$ is equal to input size, QC drops abruptly. This dramatic change is caused by the transformation of our higher radix adder from the hybrid to Gidney's RCA.

It can be seen that the performance of the higher radix adder is significantly influenced by the radix. Therefore, by changing it, we can adapt the proposed adder to the specific requirements of different scenarios as well as minimize the overall cost.
In general, when QC is more expensive, a small radix should be set to avoid introducing too much ancillas, while when T-depth or T-count is more expensive, we can try to set large radix to make T-depth or T-count small.

%% file: document/4-2-Selection.tex
The best radix for T-depth, T-count and QC is defined as the radix that leads to the lowest cost of our adder in terms of T-depth, T-count and QC, respectively.
As shown in Figure \ref{fig:fig_Compare_radix}; the fluctuations of T-depth, T-count, and QC as $r$ increases are different. Hence, the corresponding best radix may be different for T-depth, T-count, and QC.
According to the formulae for cost (given in  Equations \eqref{formula-TC}, \eqref{formula-TD} and \eqref{formula-QC}), the corresponding optimum radix can be determined. See Appendix \ref{best-r} for more details.

%% file: document/4-3-Experiment-2.tex
\begin{figure*}[!ht]
    \centering
    \includegraphics[width=\textwidth]{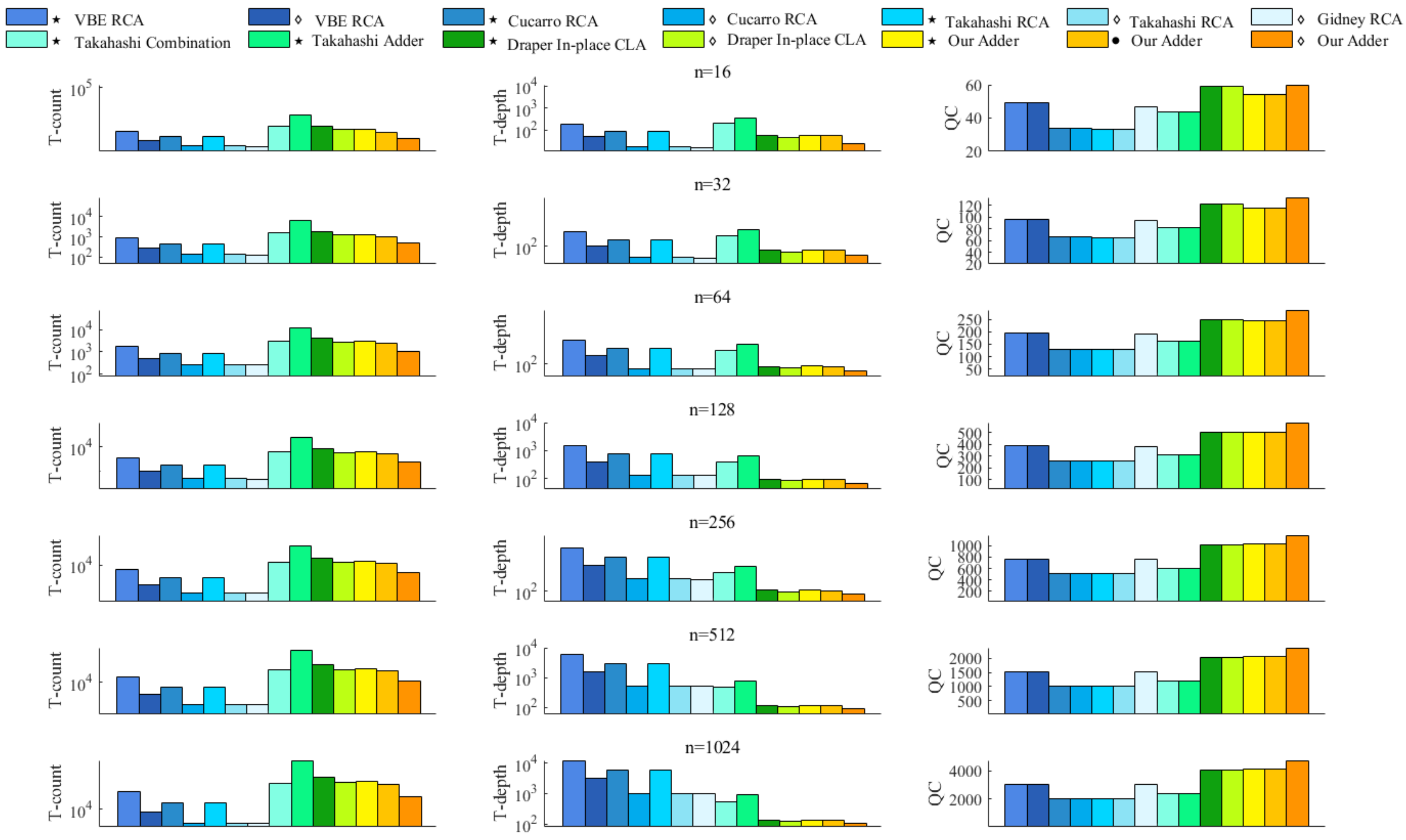}
    \caption{Comparison of the cost required by different quantum adders.\label{fig: n1024_B}}
\end{figure*}
\begin{figure*}[!ht]
    \centering
    \includegraphics[scale=0.6]{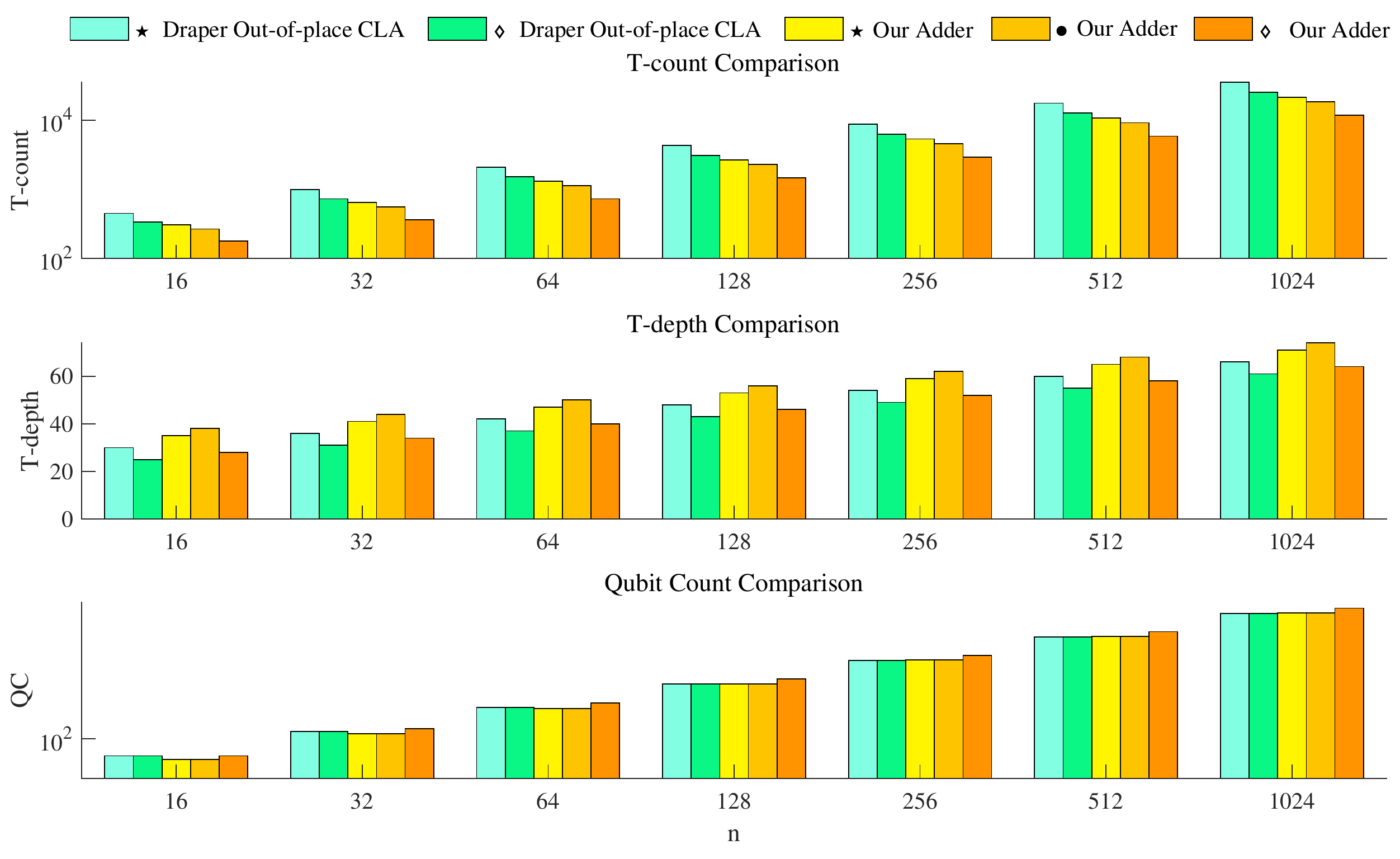}
    \caption{Comparison of the cost required by Draper's out-of-place CLAs and our adders.\label{fig: n1024_S}}
\end{figure*}

The performance of our adder compared to other well-known quantum adders is shown in the following. We summarize the cost formulas of them in Table \ref{tab:table_Compare} and \ref{tab:table_Compare_Out-of-place}. Based on them, the relevant data is visualized in Figures \ref{fig: n1024_B} and \ref{fig: n1024_S}.
 
Firstly, we describe some important experimental details. In order to evaluate our adder more objectively, three Toffoli decomposition methods are used to decompose adders into three different versions.
For adders which are denoted by \attherate, all the Toffoli pairs are decomposed using the Logical-And structure, and then the rest are decomposed using Method 3 mentioned in \secn \ref{s:adder-stru}. For adders which are denoted by \exclamation, all the Toffolis are decomposed by Method 3. For adders which are denoted by \mysignal, only Gidney's RCAs\cite{Craig_Gidney_2017} are decomposed using the Logical-And structure, and the rest are decomposed by Method 3.
Among them, the adders which are denoted by \attherate and \exclamation have smaller qubits because that Logical-And structure is not used before the sum path, which means some ancilla can be reused in sum path after the uncomputation.
For the adders which are denoted by \mysignal, the overall T-depth and T-count is reduced at the cost of QC. This is due to that it uses Logical-And structures wherever possible.

Then we compare the proposed adder with other well-known works.
Compared to quantum RCAs, our adder  consistently has a significant advantage in terms of T-depth, despite having more T-count and QC.
Compared to quantum CLAs, our adders which are denoted by \attherate and \mysignal have similar T-count and T-depth, but significantly smaller T-count. For our adder which is denoted by \exclamation, it significantly reduces the T-depth and further reduces the T-count at the cost of a slight increase in qubit.
Since Draper's out-of-place adder\cite{Draper08} does not need to be complexly uncomputed like the in-place one. Therefore, we construct a simplified version of our higher radix adder to objectively compare with. According to Fig. \ref{fig: n1024_S}, our adder slightly increases T-depth and QC, but significantly decreases T-count.
Moreover, compared to Takahashi adder \cite{Takahashi08} which is a special quantum CLA that introduces grouping idea, all the versions of the higher radix adder have similar QC and significantly smaller T-count. For T-depth, our adders which are denoted by \attherate and \mysignal are similar to Takahashi adder, but our adder which is denoted by \exclamation has a huge reduction.
Besides, the higher radix adder is also compared with Takahashi combination adder \cite{Takahashi09}, which also combines RCA and CLA. Although our QC is larger, the T-count and T-depth of our adder have different degrees of reduction. It is obvious that our structure is more general and flexible, and further improves the overall efficiency.

In general, the higher radix adder needs more qubits as a cost to significantly reduce the overall T-count and T-depth compared to other adders.

\bgroup
\begin{table*}
\renewcommand\arraystretch{1.6} 
\caption{Performance analysis of different quantum adders.\\
To simplify the representation, we compare the costs of different higher radix adders with $r$ from 3 to $n-1$ and then use the lowest cost to represent the performance of our adder. The formula for $\omega(n)$ is $\omega(n)=n-\sum_{y=1}^\infty\left \lfloor\frac{n}{2^y}\right \rfloor$ and the range for $r$ is
$2 < r \leq n$.
\label{tab:table_Compare}}
\resizebox{1.0\textwidth}{!}{
\begin{tabular}{|cc|c|c|c|}
\hline
 Adder&Year &T-count
&T-depth&QC\\ \hline

\exclamation VBE RCA \cite{VBE}&$1995$&$28n-14$&$12n-6$&$3n+1$\\\hline

\attherate VBE RCA \cite{VBE}&$1995$&$8n+6$&$3n+4$&$3n+1$\\\hline

\exclamation Cuccaro RCA \cite{Cuccaro}&$2004$&$14n-7$&$6n-3$&$2n+2$\\\hline

\attherate Cuccaro RCA \cite{Cuccaro}&$2004$&$4n+3$&$n+2$&$2n+2$\\\hline
 
\exclamation Draper In-place CLA \cite{Draper08}&$2004$& \makecell{$70n-21\omega(n)-21\omega(n-1)$\\$-21\left \lfloor\log n\right \rfloor-21\left \lfloor\log (n-1)\right \rfloor-49$ }
 &\makecell{$24+3\cdot \left \lfloor\log n\right \rfloor+3\cdot \left \lfloor\log (n-1)\right \rfloor$\\$+3\cdot \left \lfloor\log \frac{n}3\right \rfloor+3\cdot \left \lfloor\log \frac{n-1}3\right \rfloor$}&$4n-\omega(n)-\left \lfloor\log n\right \rfloor$ \\\hline

 \attherate Draper In-place CLA \cite{Draper08}&$2004$&\makecell{$50n-11\omega(n)-21\omega(n-1)$\\$-11\left \lfloor\log n\right \rfloor-21\left \lfloor\log (n-1)\right \rfloor-39$ }&
 \makecell{$15+3\cdot \left \lfloor\log n\right \rfloor+3\cdot \left \lfloor\log (n-1)\right \rfloor$\\$+3\cdot \left \lfloor\log \frac{n}3\right \rfloor+3\cdot \left \lfloor\log \frac{n-1}3\right \rfloor$}
 &$4n-\omega(n)-\left \lfloor\log n\right \rfloor$\\\hline

 \exclamation Takahashi Adder \cite{Takahashi08}&$2008$&$196n$&$90\log n$&$2n+\frac{3n}{\log n}$ \\\hline

\exclamation Takahashi RCA \cite{Takahashi09}&$2009$&$14n-7$&$6n-3$&$2n+1$\\\hline

\attherate Takahashi RCA \cite{Takahashi09}&$2009$&$4n+3$ &$n+3$&$2n+1$\\\hline
  
\exclamation Takahashi combination  \cite{Takahashi09}&$2009$&$49n$ &$54\log n$&$2n+\frac{3\cdot n}{\log n}$ \\\hline


\attherate Gidney RCA \cite{Craig_Gidney_2017}&$2018$&$4n-4$&$n$&$3n-1$\\

\hline
\rowcolor{green!6} \multicolumn{2}{|c|}{\exclamation Our Adder} &\begin{tabular}[c]{@{}c@{}}$56n-7\frac{n}{r}-7\cdot(n-1)\pmod  r$\\$-21\omega(\frac{n}{r})-21\log n +21\log r -21$\end{tabular}
&\begin{tabular}[c]{@{}c@{}}$12 \log n+9r-6\log r$\\$ -6\log 3r+6\log (r-2)+2$\end{tabular}
 &\begin{tabular}[c]{@{}c@{}}$4n-\log n+\frac{n}{r}$\\$ -\omega(\frac{n}{r})+\log r-1$\end{tabular}\\\hline
\rowcolor{yellow!6} \multicolumn{2}{|c|}{\mysignal Our Adder} &\begin{tabular}[c]{@{}c@{}}$46n+3\frac{n}{r}-7\cdot(n-1)\pmod  r $\\$ -21\omega(\frac{n}{r})-21\log n +21\log r -11$\end{tabular}
&\begin{tabular}[c]{@{}c@{}}$12 \log n+7r-6\log r$\\$ -6\log 3r+6\log (r-2)+5$\end{tabular}
 &\begin{tabular}[c]{@{}c@{}}$4n-\log n+\frac{n}{r}$\\$ -\omega(\frac{n}{r})+\log r-1$\end{tabular}\\\hline
\rowcolor{cyan!6} \multicolumn{2}{|c|}{\attherate Our Adder} & \begin{tabular}[c]{@{}c@{}}$19n+40\frac{n}{r}-7\cdot(n-1)\pmod  r $\\$ -21\omega(\frac{n}{r})-21\log n +21\log r -11$ \end{tabular}
&\begin{tabular}[c]{@{}c@{}}$12 \log n+2r-6\log r$\\$ -6\log 3r+\log (r-2)+6$\end{tabular}
 &\begin{tabular}[c]{@{}c@{}}$5n-\log n- \frac{n}{r}$\\$ -\omega(\frac{n}{r})+\log r-1$\end{tabular}\\\hline
\end{tabular}}
\end{table*}
\begin{table*}
\renewcommand\arraystretch{1.6} 
\caption{Performance analysis of quantum out-of-place CLAs.\\
Since Draper's out-of-place adder does not need to be complexly uncomputed like the in-place adder, it is unfair to compare our adder directly to it. Therefore, we construct a simplified version of our higher  radix adder to compare with.
The relevant formulas are shown below.
\label{tab:table_Compare_Out-of-place}}
\resizebox{1.0\textwidth}{!}{
\begin{tabular}{|cc|c|c|c|}
\hline
 Adder&Year&T-count
&T-depth& QC\\ \hline

  \exclamation Draper Out-of-place CLA \cite{Draper08}& $2004$ &$35n-21\omega(n)-21\left \lfloor\log n\right \rfloor-7$ &$12+3\cdot \left \lfloor\log n\right \rfloor+3\cdot \left \lfloor\log \frac{n}3\right \rfloor$&$4n+1-\omega(n)-\left \lfloor\log n\right \rfloor$ \\\hline
 
 \attherate Draper Out-of-place CLA \cite{Draper08}&$2004$&$25n-11\omega(n)-11\left \lfloor\log n\right \rfloor-7$&
 $7+3\cdot \left \lfloor\log n\right \rfloor+3\cdot \left \lfloor\log \frac{n}3\right \rfloor$
 &$4n+1-\omega(n)-\left \lfloor\log n\right \rfloor$\\

\hline
 \rowcolor{green!6} \multicolumn{2}{|c|}{\exclamation Our Adder} &\begin{tabular}[c]{@{}c@{}}$35n-7\cdot(n-1)\pmod  r+7\cdot n/r
$\\$ 
-14\omega(\frac{n}{r})-14\log n +14\log r -14$\end{tabular}
 &\begin{tabular}[c]{@{}c@{}}$6 \log n+3r-3\log r$\\$ -3\log 3r+3\log (r-2)+5$\end{tabular}
 &\begin{tabular}[c]{@{}c@{}}$4n-\log n+\frac{n}{r}$\\$ -\omega(\frac{n}{r})+\log r-1$\end{tabular}\\\hline
  \rowcolor{yellow!6}  \multicolumn{2}{|c|}{\mysignal Our Adder} &\begin{tabular}[c]{@{}c@{}}$32n+10\frac{n}{r}-7\cdot(n-1)\pmod  r $\\$ -14\omega(\frac{n}{r})-14\log n +14\log r -11$\end{tabular}
&\begin{tabular}[c]{@{}c@{}}$6 \log n+4r-3\log r$\\$ -3\log 3r+3\log (r-2)+5$\end{tabular}
 &\begin{tabular}[c]{@{}c@{}}$4n-\log n+\frac{n}{r}$\\$ -\omega(\frac{n}{r})+\log r-1$\end{tabular}\\\hline
\rowcolor{cyan!6} \multicolumn{2}{|c|}{\attherate Our Adder} &\begin{tabular}[c]{@{}c@{}}$19n+26\frac{n}{r}-7\cdot(n-1)\pmod  r $\\$ -14\omega(\frac{n}{r})-14\log n +14\log r -11$\end{tabular}
&\begin{tabular}[c]{@{}c@{}}$6 \log n+2r-3\log r$\\$ -3\log 3r+\log (r-2)+3$\end{tabular}
 &\begin{tabular}[c]{@{}c@{}}$5n-\log n- \frac{n}{r}$\\$ -\omega(\frac{n}{r})+\log r-1$\end{tabular}\\\hline
 
\end{tabular}}
\end{table*}
\egroup

%% file: document/4-4-Connecting.tex
\begin{figure}[!ht]
    \centering
     \includegraphics[width=.6\linewidth]{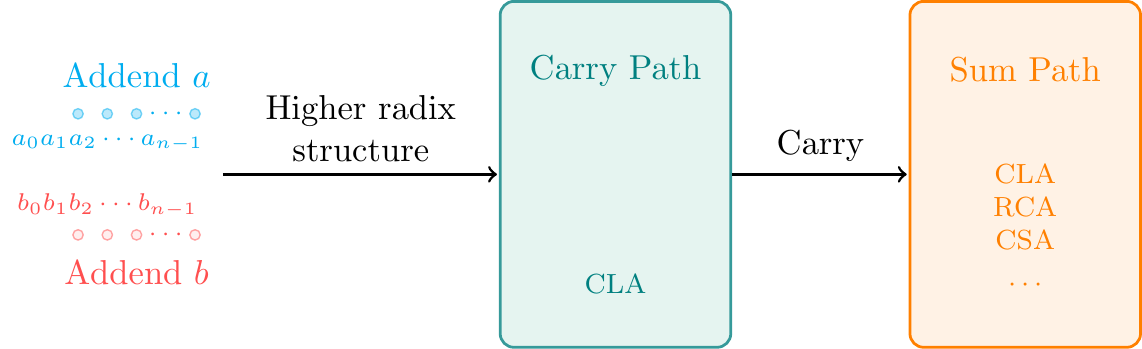}
    \caption{A general framework of higher radix adder. \label{fig:Workflow}}
\end{figure}

Our work can be seen as a bridge to connect existing quantum adders. Figure \ref{fig:Workflow} illustrates the general framework of it, whose key parts are the carry path and the sum path.

\begin{figure}[!ht]
    \centering
    \subfigure[Visualization.]{\includegraphics[width=.46\linewidth]{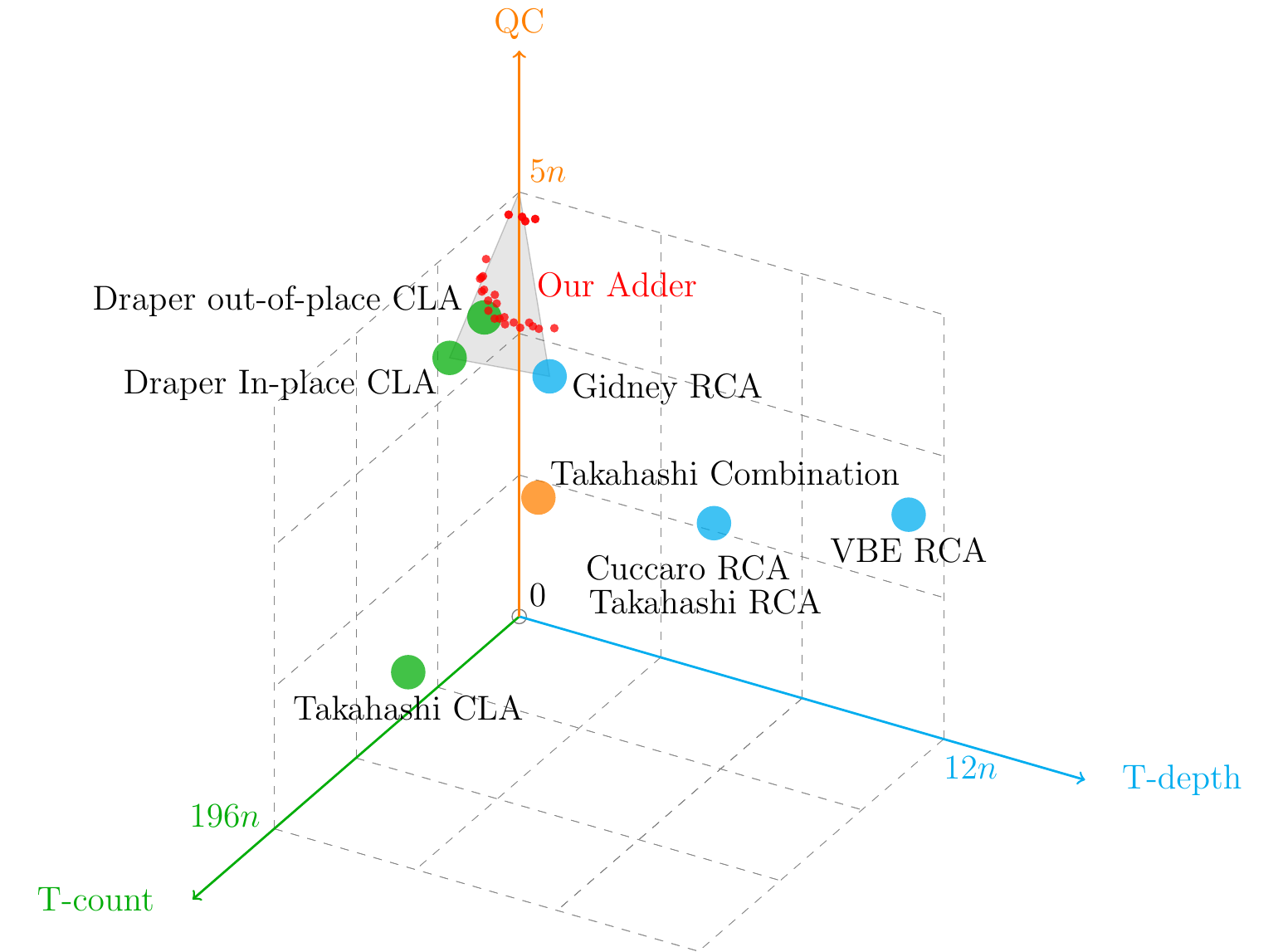}}
    \subfigure[Zoom in on the gray scope.]
    {\includegraphics[width=.46\linewidth]{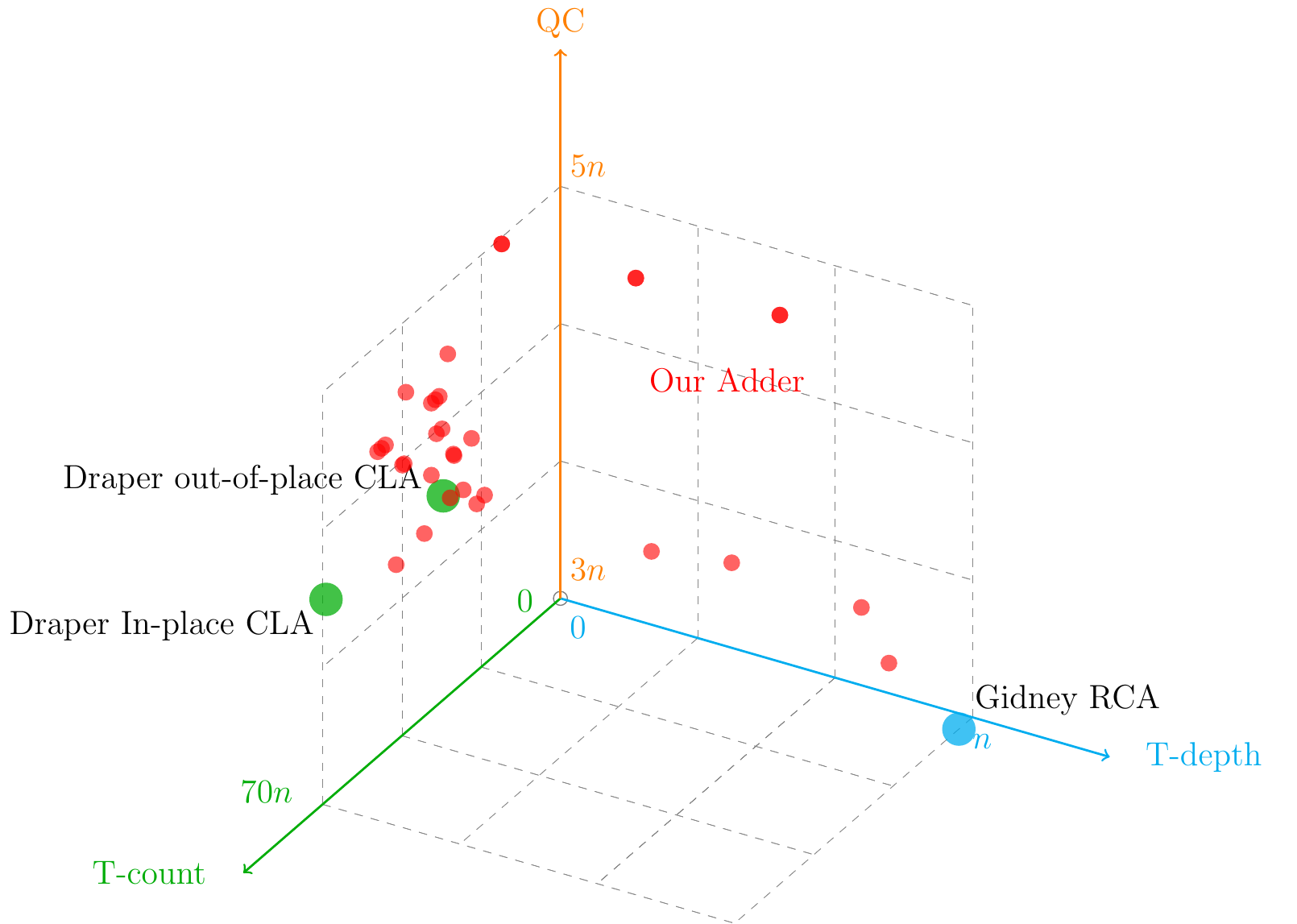}}
    \caption{Visualization of quantum adders. \label{f:3d}}
\end{figure}
For the carry path, quantum CLAs can be used to compute specific carries.
For the sum path, any quantum adder can be used to calculate the final result based on those carries.
It is interesting to note that the cost contribution of the carry path and sum path in the total circuit is adjusted by changing radix. More specifically, when the radix is large, the number of groups which is divided by the higher radix layer is small. Hence, the carry chain is short, which means the sum path is a larger cost contribution of the overall circuit than the carry path. On the contrary, when the radix is small, the carry chain is longer, which means the carry path accounts for a large portion of the total circuit.

According to this general framework, this work can be seen as a specific example based on Draper's CLA and Gidney's RCA. Specifically, our carry path uses the same Brent-Kung tree structure as Draper's CLA, and our sum path is Gidney's RCA. Apart from these two adders, other quantum adders can also be used to construct a higher radix adder.

In order to support one to construct the cheapest quantum adder in different scenarios quickly and easily, we summarize the performance of well-known quantum adders in Fig. \ref{f:3d}.
When QC is more expensive, it is more suitable to use adders with less ancilla such as Takahashi RCA. 
For T-count, RCAs such as Gidney's RCA can effectively reduce the overall cost.
For reducing T-depth, it is recommended to integrate Draper's In-place CLA or other quantum CLAs within higher radix framework described in Fig.\ref{fig:Workflow}.

%% file: document/5-conclusion.tex
Quantum adder is one of the most fundamental components in quantum computing. Therefore, designing a quantum adder with lower cost is of great significance for establishing a more efficient and cheaper large-scale quantum circuit.
This paper proposed an efficient quantum circuit for integer addition by introducing techniques from classical higher radix carry-lookahead adder and Manchester Carry Chain adder.
In terms of T-depth and T-count, the proposed circuit is superior to all the existing quantum carry-lookahead adders except Draper Out-of-place CLA. Compared with Draper's Out-of-place CLA, the proposed higher radix adder has significantly lower T-count with comparable QC and T-depth.
Due to practical constraints, we only analyzed three main quantum circuit complexity metrics, T-count, T-depth, and QC. 

In the future, one may be interested in how to automatically design the best adder based on specific cost constraints and how to accurately and quickly tune the radix to obtain the most efficient adder. Additionally, exploring the T-count, T-depth and QC limits of quantum addition is also a meaningful and challenging problem. Finally, comparing various adder designs considering practical constraints, such as quantum error correction (QEC) and  topological structures is an important open problem.

%% file: document/6-appendix.tex
Here, we use several specific numerical examples to show the workflow of the higher radix adder. 

We assume that the binary number $a$ is $101001$, and $b$ is $010011$. Additionally, the first carry bit $c_0$ is set to $0$. Our target is to get the sum $s$ of $a$ and $b$.
For this specific six-bit addition, when the radix is changed from from $6$ to $1$, the calculation process is as follows.

\begin{itemize}
    \item {$\mathbf{r=6.}$}
    
When the radix is equal to the bit width of the binary addends, our higher radix adder is just a simple ripple carry adder that only calculates the sum path. The calculation process of the sum path is as follows.
    
\qquad \textcolor{blue}{Sum path:}
\begin{tabular}{ccc} 
\centering
 &$a$&$101001$\\
+&$b$&$010011$\\
\hline
 &$s$&$111100$\\
\end{tabular}\qquad
\textcolor{blue}{Calculate $p$:}
\begin{tabular}{ccc} 
\centering
 &$a$&$101001$\\
$\oplus$&$b$&$010011$\\
\hline
 &$p$&$111010$\\
\end{tabular}\qquad
\textcolor{blue}{Calculate $g$:}
\begin{tabular}{ccc} 
\centering
 &$a$&$101001$\\
$\times$&$b$&$010011$\\
\hline
 &$g$&$000001$\\
\end{tabular}
\end{itemize}

For the remaining adders, most of them need to go through four steps, $p$ and $g$ calculation, higher radix structure, carry path, and sum path.
The first step, calculating $p$ and $g$, is the same for all the remaining examples. The calculation process is shown above.

After obtaining $p$ and $g$, the remaining steps differ for the higher radix adder with different radix. The details of the calculation of our higher radix adder with its radix set from $5$ to $1$ are shown below.

\begin{itemize}
    \item {$\mathbf{r=5.}$} To begin with, since we do not need to compute the carry of the most significant bit, the most significant $p$ and $g$ are ignored.
Firstly, the higher radix structure divides the remaining $p$ and $g$ into groups of five each, yielding a $p_{group}$ and a $g_{group}$.
In the carry path, according to the Brent-Kung structure, we do not need any operation. In this case, $c_5$ and $g_{group}$ are equal, both being $0$.
In the sum path, the ripple carry structure is used to add $a$, $b$; and the computed carry $c$ by the group to obtain the sum $s$.

\textcolor{blue}{Higher radix structure:\qquad\qquad \qquad\qquad \qquad\qquad\qquad\qquad\qquad\quad\qquad\qquad Carry path:\qquad\qquad\qquad}
\\
\includegraphics[scale=0.88]{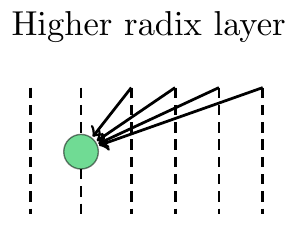}\qquad
\includegraphics[scale=0.88]{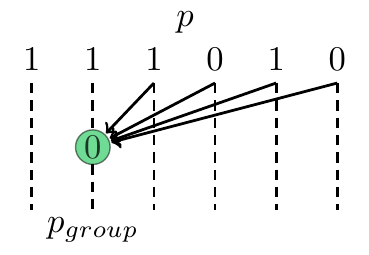}\qquad
\includegraphics[scale=0.88]{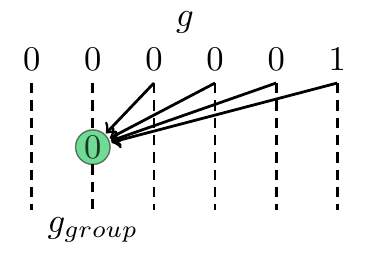}\qquad
\includegraphics[scale=.88]
{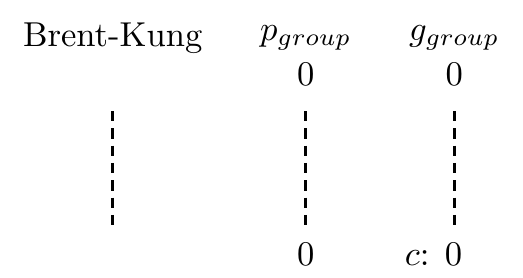}
\quad \\
 \textcolor{blue}{Sum path:}
\begin{tabular}{ccc}
 &$a$&$101001$\\
 &$b$&$010011$\\
 &$c$&$0\quad\quad0$\\
\hline
 &$s$&$111100$\\
\end{tabular}
    \item {$\mathbf{r=4.}$} Similar to radix $5$, we show radix-$4$ addition below.
    
\textcolor{blue}{Higher radix structure:\qquad\qquad \qquad\qquad \qquad\qquad\qquad\qquad\qquad\quad\qquad\qquad Carry path:}

\includegraphics[scale=0.88]{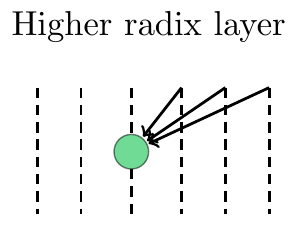}\qquad
\includegraphics[scale=0.88]{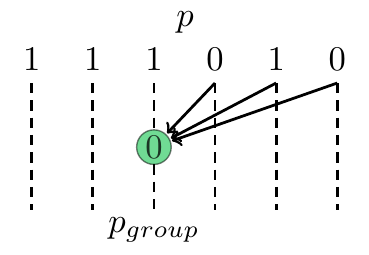}\qquad
\includegraphics[scale=0.88]{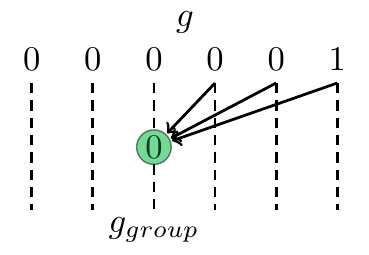}\qquad
\includegraphics[scale=.88]
{Diagrams/Appendix-A/Carry_5_4_3.pdf}\qquad

\textcolor{blue}{Sum path:}
\begin{tabular}{ccc} 
\centering
 &$a$&$101001$\\
 &$b$&$010011$\\
 &$c$&$\,\,\,\,\,0\,\,\,\,\quad 0$\\
\hline
 &$s$&$111100$\\
\end{tabular}
    \item {$\mathbf{r=3.}$} Similarly, the following is the process for an adder with radix $3$.

\textcolor{blue}{Higher radix structure:\qquad\qquad \qquad\qquad \qquad\qquad\qquad\qquad\qquad\quad\qquad\qquad Carry path:}

\includegraphics[scale=0.88]{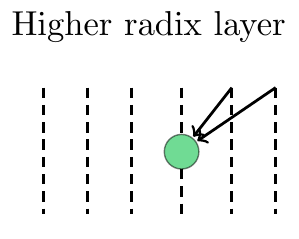}\qquad
\includegraphics[scale=0.88]{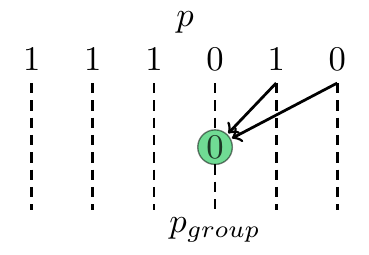}\qquad
\includegraphics[scale=0.88]{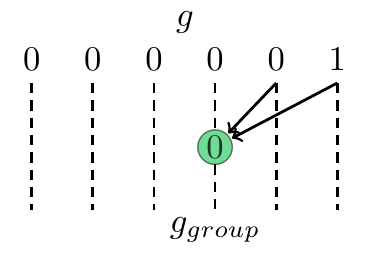}\qquad
\includegraphics[scale=.88]
{Diagrams/Appendix-A/Carry_5_4_3.pdf}\qquad

\textcolor{blue}{Sum path:}
\begin{tabular}{ccc} 
\centering
 &$a$&$101001$\\
 &$b$&$010011$\\
 &$c$&$\,\,\,\,\,\,\,\,\,0\,\,\,\,\,\,\,\,\,0$\\
\hline
 &$s$&$111100$\\
\end{tabular}
    \item {$\mathbf{r=2.}$} The overall calculation process is shown as follows.\\
\textcolor{blue}{Higher radix structure:\qquad\qquad \qquad\qquad \qquad\qquad\qquad\qquad\qquad\quad\qquad\qquad Carry path:}

\includegraphics[scale=0.88]{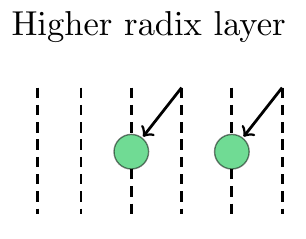}\qquad
\includegraphics[scale=0.88]{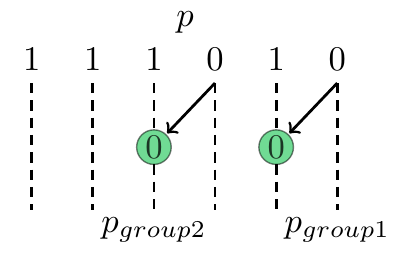}\qquad
\includegraphics[scale=0.88]{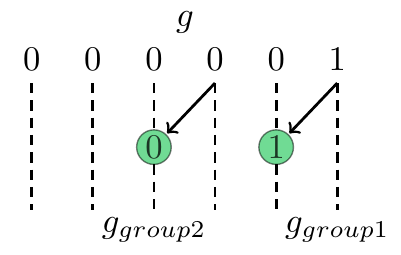}\qquad
\includegraphics[scale=.88]
{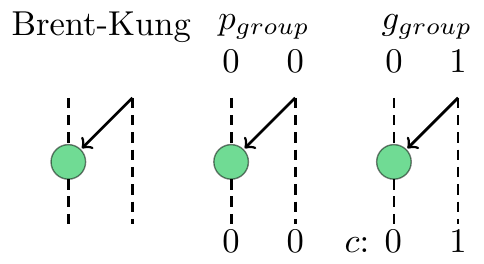}\qquad

\textcolor{blue}{Sum path:}
\begin{tabular}{ccc} 
\centering
 &$a$&$101001$\\
 &$b$&$010011$\\
 &$c$&$\,\,\,\,\,0\,\,\,\,1\,\,\,\,0$\\
\hline
 &$s$&$111100$\\
\end{tabular}
    \item {$\mathbf{r=1.}$} When the radix is one, it is a special case. Since the higher radix structure does not work, it is essentially a CLA. 

\textcolor{blue}{Carry path}:\\
\includegraphics[scale=.88]
{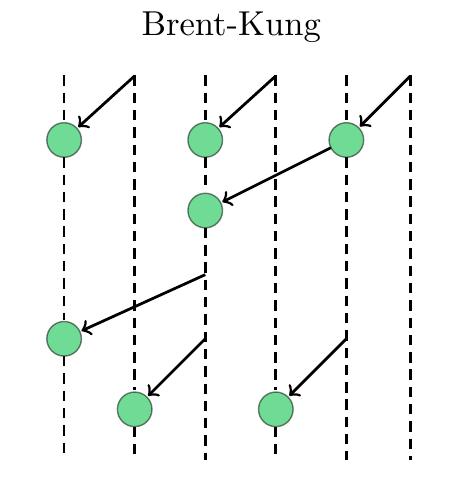}
\includegraphics[scale=.88]
{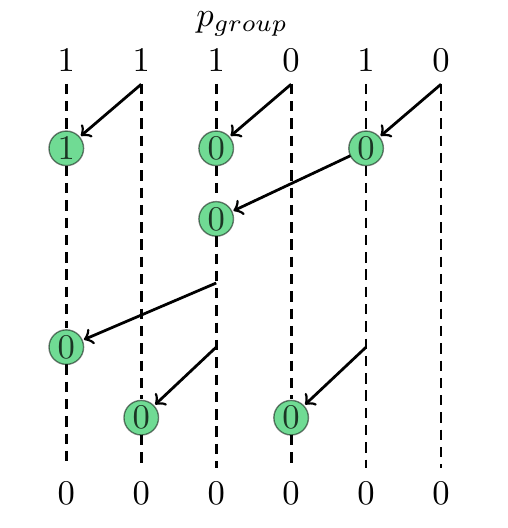}
\includegraphics[scale=.88]
{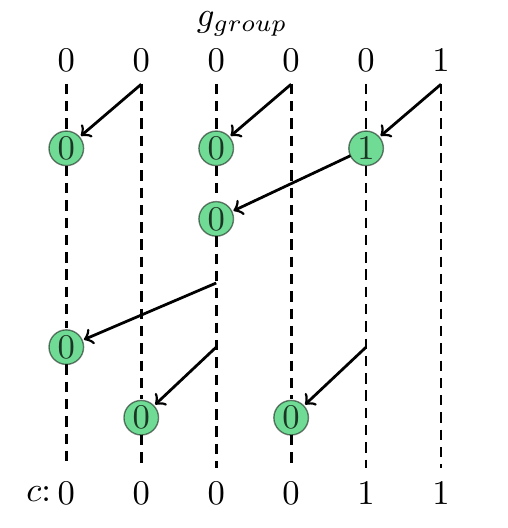}

\textcolor{blue}{Sum path:}
\begin{tabular}{ccc} 
\centering
 &$a$&$101001$\\
 &$b$&$010011$\\
 &$c$&$000110$\\
\hline
 &$s$&$111100$\\
\end{tabular}
\end{itemize}

%% file: document/7-appendix.tex
Here, we show the details of adding the sub-cost formulae  of the seven steps to obtain the final cost formulae  in section \ref{ss:The whole circuit}.
%
In Derivations \eqref{deri:t-count}, \eqref{deri:t-depth}, \eqref{deri:qc} respectively; we show the step-by-step derivation for the T-count and T-depth.

\begin{equation}\label{deri:t-count}
\begin{split}
\allowdisplaybreaks
    &\text{T-count}\\
    =&\overbrace{7(n-\alpha)}^{\text{Step 1}}+\overbrace{\rho [7+4(2 r-3)] }^{\text{Step 2}}
    +\overbrace{2\cdot 7 [2 \rho -1 -\omega (\rho)
    -\lfloor\log (\rho)\rfloor] }^{\text{Step 3}}
    +\overbrace{7 [2 \rho -1 -\omega (\rho)-\lfloor\log (\rho)\rfloor]}^{\text{Step 4}}
    +\overbrace{0 }^{\text{Step 5}}
    +\overbrace{7\rho}^{\text{Step 6}}+\overbrace{4(n-\lceil\frac{n}{r} \rceil)}^{\text{Step 7}}\\
    =& 7 n-7 \alpha +7 \rho +8 r\rho-12 \rho
    +28 \rho-14 -14\omega (\rho)+14\rho
    -14\lfloor\log (\rho)\rfloor-7-7\omega (\rho)
    -7\lfloor\log (\rho)\rfloor +7 \rho+4n -4\lceil\frac{n}{r} \rceil \\
    =& 44\rho -4(\lceil\frac{n}{r} \rceil-1)-4+8 r\rho 
     +11 n-7 \alpha -21-21\omega (\rho) -21\lfloor\log (\rho)\rfloor \\
    =& 40\rho +8r\rho+11 n -7 \alpha -25
    -21\omega (\rho)-21\lfloor\log (\rho)\rfloor\\
    =& (8r+40)(\lceil\frac{n}{r} \rceil-1)+11 n -25  
    -7 [(n-1)\pmod r +1] 
    -21\omega (\lceil\frac{n}{r} \rceil-1)
    -21\lfloor\log (\lceil\frac{n}{r} \rceil-1)\rfloor
    \\
    =& (8r+40)(\lceil\frac{n}{r} \rceil)+11 n -7-25
    -40 -8r -7 \cdot(n-1)\pmod r
    -21\omega (\lceil\frac{n}{r} \rceil-1)
    -21\lfloor\log (\lceil\frac{n}{r} \rceil-1)\rfloor 
    \\=&(8r+40)\cdot\left \lceil\frac{n}{r}\right \rceil+11n 
    -72 -8r 
    -7\cdot(n-1)\pmod r
    -21\omega(\left \lceil\frac{n}{r}\right \rceil-1) 
    -21\left \lfloor\log (\left \lceil\frac{n}{r}\right \rceil-1) \right \rfloor)\\
\end{split}
\end{equation}

\begin{equation}\label{deri:t-depth}
\begin{split}
\allowdisplaybreaks
&\text{T-depth}\\
=& \overbrace{3}^{\text{Step 1}}+\overbrace{[3+r-1+\beta]}^{\text{Step 2}}
+\overbrace{3\cdot( \left \lfloor\log (\rho)\right \rfloor+ \left \lfloor\log \frac{\rho}3\right \rfloor+2)}^{\text{Step 3}}
+\overbrace{3\cdot( \left \lfloor\log \rho\right \rfloor+ \left \lfloor\log \frac{\rho}3\right \rfloor+1)}^{\text{Step 4}}
+\overbrace{0}^{\text{Step 5}}+\overbrace{3}^{\text{Step 6}}+\overbrace{r}^{\text{Step 7}} \\
=& 6( \left \lfloor\log (\rho)\right \rfloor+ \left \lfloor\log \frac{\rho}3\right \rfloor)+17+2r+\beta \\
=& 6( \left \lfloor\log (\rho)\right \rfloor+ \left \lfloor\log \frac{\rho}3\right \rfloor+4)+17
+2-24+\left \lfloor\log(r-2) \right \rfloor +2r
    \\=&6\cdot (\left \lfloor\log (\left \lceil\frac{n}{r}\right \rceil -1) \right \rfloor
    +\left \lfloor\log (\frac{1}{3}(\left \lceil\frac{n}{r}\right \rceil -1)) \right \rfloor+4)
    +\left \lfloor\log(r-2) \right \rfloor+2r-5
\end{split}
\end{equation}

\begin{equation}\label{deri:qc}
\begin{split}
\allowdisplaybreaks
 &\text{QC}\\
 =& \overbrace{[3\cdot n-\alpha]}^{\text{Step 1}}+
 \overbrace{[\rho\cdot (r-1)]}^{\text{Step 2}}
 +\overbrace{[2\cdot \rho -1 - \omega(\rho)-\lfloor\log (\rho)\rfloor]}^{\text{Step 3}}
 +\overbrace{0}^{\text{Step 4}}+
 \overbrace{0}^{\text{Step 5}}+
 \overbrace{0}^{\text{Step 6}}+
 \overbrace{[\alpha-1+(r-2)\rho]}^{\text{Step 7}} \\
=& 3n-\alpha+r\rho-\rho +2\rho-1
-\omega (\rho)-\lfloor\log (\rho)\rfloor+\alpha-1
+r\rho-2\rho  \\
=& 3n+(2r-1)\rho-2-\omega (\rho)
-\lfloor\log (\rho)\rfloor \\
=& 3n+(2r-1)(\lceil\frac{n}{r} \rceil-1)-2
-\omega (\lceil\frac{n}{r} \rceil-1) 
-\lfloor\log (\lceil\frac{n}{r} \rceil-1)\rfloor \\
=& 3n+1-2-2r-\omega (\lceil\frac{n}{r} \rceil-1)
+(2r-1)\lceil\frac{n}{r} \rceil-\lfloor\log (\lceil\frac{n}{r} \rceil-1)\rfloor  
    \\=&3n-1-2r-\omega(\left \lceil\frac{n}{r}\right \rceil-1)
    +(2r-1)\cdot\left \lceil\frac{n}{r}\right \rceil
    -\left \lfloor\log (\left \lceil\frac{n}{r}\right \rceil-1) \right \rfloor\\
 \end{split}
\end{equation}

%% file: document/8-appendix.tex
According to the cost formulae in Table \ref{tab:table_Compare}, here we will analyze how to find the best radix in the case of large input size in detail.
First, we divide the radix into three cases: small, medium and large, and use $r_s$, $r_m$ and $r_l$ to denote them respectively. Furthermore, $n$ is used to denote input size and $r$ is used to denote the radix. 
It is assumed that $O(r)$ is equal to $O(1)$ in the $r_s$ case, $O(r)$ is equal to $O(\sqrt{n})$ in the $r_m$ case, and $O(r)$ is equal to $O(n)$ in the $r_l$ case. Besides, $r$ should be strictly less than or equal to $n$.

For our adder which is denoted by \exclamation, the asymptotic complexity of T-count, T-depth and QC is shown below.
 \begin{align}
r_s &
 \begin{cases}
        T-count&=56n-7\frac{n}{r}; \\
        T-depth&=12\log n;  \\
        QC&=4n+\frac{n}{r} \\
\end{cases}
r_m &
 \begin{cases}
        T-count&=56n ;\\
        T-depth&=12 \log n+9r ;\\
        QC&=4n \\
\end{cases}
 r_l &
 \begin{cases}
        T-count&=56n-7(n-1)\pmod r;\\
        \textcolor{gray}{T-count}& \textcolor{gray}{\geq 56n-7r;}\\
        T-depth&=9r;\\
        QC&=4n\\
\end{cases}
 \end{align}
For T-depth, the best radix is in the range of $r_s$ or $r_l$. For T-count, the best radix is in the $r_s$ range. For QC, the best radix is in the range of $r_l$.

For our adder which is denoted by \mysignal, the asymptotic complexity of T-count, T-depth and QC is shown below.
 \begin{align}
r_s &
 \begin{cases}
        T-count&=46n+3\frac{n}{r}; \\
        T-depth&=12\log n ;  \\
        QC&= 4n+\frac{n}{r}\\
\end{cases}
r_m &
 \begin{cases}
        T-count&=46n;\\
        T-depth&=12 \log n+7r;\\
        QC&= 4n\\
\end{cases}
 r_l &
 \begin{cases}
        T-count&=46n-7(n-1)\pmod  r;\\
        \textcolor{gray}{T-count}& \textcolor{gray}{\geq 46n-7r;}\\
        T-depth&=7r;\\
        QC&=4n\\
\end{cases}
 \end{align}
For T-depth, the best radix is in the range of $r_l$. For T-count, the best radix is in the $r_s$ range. For QC, the best radix is in the range of $r_l$.

For our adder which is denoted by \attherate, the asymptotic complexity of T-count, T-depth and QC is shown below.
 \begin{align}
r_s &
 \begin{cases}
        T-count&=19n+40\frac{n}{r}; \\
        T-depth&=12\log n ;  \\
        QC&=5n-\frac{n}{r} \\
\end{cases}
r_m &
 \begin{cases}
        T-count&=19n;\\
        T-depth&=12 \log n+2r;\\
        QC&=5n \\
\end{cases}
 r_l &
 \begin{cases}
        T-count&=19n-7(n-1)\pmod  r;\\
        \textcolor{gray}{T-count}& \textcolor{gray}{\geq 19n-7r;}\\
        T-depth&=2r;\\
        QC&=5n\\
\end{cases}
 \end{align}
For T-depth, the best radix is in the range of $r_l$. For T-count, the best radix is in the $r_s$ range. For QC, the best radix is in the range of $r_s$.

%% file: document/10-appendix.tex
\setcounter{table}{0}
\thispagestyle{empty}

\section*{All tables in this paper}
\begin{table}[!ht]
\caption{\label{table_Toffoli_decomposition}%
Summary table of Toffoli decomposition.
}
\centering
\begin{tabular}{|c|c|c|c|}
\hline
\textrm{Decomposition}&
\textrm{T-count}&
\textrm{T-depth}&
\textrm{Ancilla Count}\\
\hline
Method 1 (Figure \ref{fig: Decomposition_Toffoli-a}) & 7 & \textcolor[rgb]{1,0,0}{6} & \textcolor[rgb]{0,0,1}{0}\\\hline
Method 2 (Figure \ref{fig: Decomposition_Toffoli-b})& 7 & 4 & \textcolor[rgb]{0,0,1}{0}\\\hline
Method 3 (Figure \ref{fig: Method 3})& 7 & 3 & \textcolor[rgb]{0,0,1}{0}\\\hline
Method 4 (Figure \ref{fig: Decomposition_Toffoli-d})& 7 & 2 & 1\\\hline
Method 5 (Figure \ref{fig: Decomposition_Toffoli-e})& 7 &\textcolor[rgb]{0,0,1}{1} & \textcolor[rgb]{1,0,0}{4}\\\hline
\end{tabular}
\end{table}

\begin{table}[!ht]
\caption{\label{table_Toffoli_decomposition}%
Summary table of Toffoli decomposition.
}
\centering
\begin{tabular}{|c|c|c|c|}
\hline
\textrm{Decomposition}&
\textrm{T-count}&
\textrm{T-depth}&
\textrm{Ancilla Count}\\
\hline
Method 1 & 7 & \textcolor[rgb]{1,0,0}{6} & \textcolor[rgb]{0,0,1}{0}\\\hline
Method 2 & 7 & 4 & \textcolor[rgb]{0,0,1}{0}\\\hline
Method 3 & 7 & 3 & \textcolor[rgb]{0,0,1}{0}\\\hline
Method 4 & 7 & 2 & 1\\\hline
Method 5 & 7 &\textcolor[rgb]{0,0,1}{1} & \textcolor[rgb]{1,0,0}{4}\\\hline
\end{tabular}
\end{table}


\begin{table}[!ht]
\caption{Different structures of sum path ($r=n$). \\When calculating T-depth, we assume that Part 1 of the CSA has been completed when calculating the carry path. Therefore, for CSA1 and CSA2, T-depth equals to the T-depth of Part 2, which is the minimum T-depth of sum path for CSAs.\label{tab:sum_path}}
\centering
\begin{tabular}{|c|c|c|c|}\hline
\textrm{Structure}&
\textrm{T-count}&
\textrm{T-depth}&
\textrm{QC}\\
\hline
CSA1 & $11n-4$ & $4n$ & $6n+1$\\\hline
CSA2 & $11n-4$ & $3n$ & $6n+1$\\\hline
RCA  & \textcolor[rgb]{0,0,1}{$4n-4$} & \textcolor[rgb]{0,0,1}{$n$} & \textcolor[rgb]{0,0,1}{$3n$}\\
\hline
\end{tabular}
\end{table}


\bgroup
\begin{table*}
\renewcommand\arraystretch{1.6} 
\caption{Performance analysis of different quantum adders.\\
To simplify the representation, we compare the costs of different higher radix adders with $r$ from 3 to $n-1$ and then use the lowest cost to represent the performance of our adder. The formula for $\omega(n)$ is $\omega(n)=n-\sum_{y=1}^\infty\left \lfloor\frac{n}{2^y}\right \rfloor$ and the range for $r$ is
$2 < r \leq n$.
\label{tab:table_Compare}}
\resizebox{1.0\textwidth}{!}{
\begin{tabular}{|cc|c|c|c|}
\hline
 Adder&Year &T-count
&T-depth&QC\\ \hline

\exclamation VBE RCA \cite{VBE}&$1995$&$28n-14$&$12n-6$&$3n+1$\\\hline

\attherate VBE RCA \cite{VBE}&$1995$&$8n+6$&$3n+4$&$3n+1$\\\hline

\exclamation Cuccaro RCA \cite{Cuccaro}&$2004$&$14n-7$&$6n-3$&$2n+2$\\\hline

\attherate Cuccaro RCA \cite{Cuccaro}&$2004$&$4n+3$&$n+2$&$2n+2$\\\hline
 
\exclamation Draper In-place CLA \cite{Draper08}&$2004$& \makecell{$70n-21\omega(n)-21\omega(n-1)$\\$-21\left \lfloor\log n\right \rfloor-21\left \lfloor\log (n-1)\right \rfloor-49$ }
 &\makecell{$24+3\cdot \left \lfloor\log n\right \rfloor+3\cdot \left \lfloor\log (n-1)\right \rfloor$\\$+3\cdot \left \lfloor\log \frac{n}3\right \rfloor+3\cdot \left \lfloor\log \frac{n-1}3\right \rfloor$}&$4n-\omega(n)-\left \lfloor\log n\right \rfloor$ \\\hline

 \attherate Draper In-place CLA \cite{Draper08}&$2004$&\makecell{$50n-11\omega(n)-21\omega(n-1)$\\$-11\left \lfloor\log n\right \rfloor-21\left \lfloor\log (n-1)\right \rfloor-39$ }&
 \makecell{$15+3\cdot \left \lfloor\log n\right \rfloor+3\cdot \left \lfloor\log (n-1)\right \rfloor$\\$+3\cdot \left \lfloor\log \frac{n}3\right \rfloor+3\cdot \left \lfloor\log \frac{n-1}3\right \rfloor$}
 &$4n-\omega(n)-\left \lfloor\log n\right \rfloor$\\\hline

 \exclamation Takahashi Adder \cite{Takahashi08}&$2008$&$196n$&$90\log n$&$2n+\frac{3n}{\log n}$ \\\hline

\exclamation Takahashi RCA \cite{Takahashi09}&$2009$&$14n-7$&$6n-3$&$2n+1$\\\hline

\attherate Takahashi RCA \cite{Takahashi09}&$2009$&$4n+3$ &$n+3$&$2n+1$\\\hline
  
\exclamation Takahashi combination  \cite{Takahashi09}&$2009$&$49n$ &$54\log n$&$2n+\frac{3\cdot n}{\log n}$ \\\hline


\attherate Gidney RCA \cite{Craig_Gidney_2017}&$2018$&$4n-4$&$n$&$3n-1$\\

\hline
\rowcolor{green!6} \multicolumn{2}{|c|}{\exclamation Our Adder} &\begin{tabular}[c]{@{}c@{}}$56n-7\frac{n}{r}-7\cdot(n-1)\pmod  r$\\$-21\omega(\frac{n}{r})-21\log n +21\log r -21$\end{tabular}
&\begin{tabular}[c]{@{}c@{}}$12 \log n+9r-6\log r$\\$ -6\log 3r+6\log (r-2)+2$\end{tabular}
 &\begin{tabular}[c]{@{}c@{}}$4n-\log n+\frac{n}{r}$\\$ -\omega(\frac{n}{r})+\log r-1$\end{tabular}\\\hline
\rowcolor{yellow!6} \multicolumn{2}{|c|}{\mysignal Our Adder} &\begin{tabular}[c]{@{}c@{}}$46n+3\frac{n}{r}-7\cdot(n-1)\pmod  r $\\$ -21\omega(\frac{n}{r})-21\log n +21\log r -11$\end{tabular}
&\begin{tabular}[c]{@{}c@{}}$12 \log n+7r-6\log r$\\$ -6\log 3r+6\log (r-2)+5$\end{tabular}
 &\begin{tabular}[c]{@{}c@{}}$4n-\log n+\frac{n}{r}$\\$ -\omega(\frac{n}{r})+\log r-1$\end{tabular}\\\hline
\rowcolor{cyan!6} \multicolumn{2}{|c|}{\attherate Our Adder} & \begin{tabular}[c]{@{}c@{}}$19n+40\frac{n}{r}-7\cdot(n-1)\pmod  r $\\$ -21\omega(\frac{n}{r})-21\log n +21\log r -11$ \end{tabular}
&\begin{tabular}[c]{@{}c@{}}$12 \log n+2r-6\log r$\\$ -6\log 3r+\log (r-2)+6$\end{tabular}
 &\begin{tabular}[c]{@{}c@{}}$5n-\log n- \frac{n}{r}$\\$ -\omega(\frac{n}{r})+\log r-1$\end{tabular}\\\hline
\end{tabular}}
\end{table*}
\begin{table*}
\renewcommand\arraystretch{1.6} 
\caption{Performance analysis of quantum out-of-place CLAs.\\
Since Draper's out-of-place adder does not need to be complexly uncomputed like the in-place adder, it is unfair to compare our adder directly to it. Therefore, we construct a simplified version of our higher  radix adder to compare with.
The relevant formulas are shown below.
\label{tab:table_Compare_Out-of-place}}
\resizebox{1.0\textwidth}{!}{
\begin{tabular}{|cc|c|c|c|}
\hline
 Adder&Year&T-count
&T-depth& QC\\ \hline

  \exclamation Draper Out-of-place CLA \cite{Draper08}& $2004$ &$35n-21\omega(n)-21\left \lfloor\log n\right \rfloor-7$ &$12+3\cdot \left \lfloor\log n\right \rfloor+3\cdot \left \lfloor\log \frac{n}3\right \rfloor$&$4n+1-\omega(n)-\left \lfloor\log n\right \rfloor$ \\\hline
 
 \attherate Draper Out-of-place CLA \cite{Draper08}&$2004$&$25n-11\omega(n)-11\left \lfloor\log n\right \rfloor-7$&
 $7+3\cdot \left \lfloor\log n\right \rfloor+3\cdot \left \lfloor\log \frac{n}3\right \rfloor$
 &$4n+1-\omega(n)-\left \lfloor\log n\right \rfloor$\\

\hline
 \rowcolor{green!6} \multicolumn{2}{|c|}{\exclamation Our Adder} &\begin{tabular}[c]{@{}c@{}}$35n-7\cdot(n-1)\pmod  r+7*n/r
$\\$ 
-14\omega(\frac{n}{r})-14\log n +14\log r -14$\end{tabular}
 &\begin{tabular}[c]{@{}c@{}}$6 \log n+3r-3\log r$\\$ -3\log 3r+3\log (r-2)+5$\end{tabular}
 &\begin{tabular}[c]{@{}c@{}}$4n-\log n+\frac{n}{r}$\\$ -\omega(\frac{n}{r})+\log r-1$\end{tabular}\\\hline
  \rowcolor{yellow!6}  \multicolumn{2}{|c|}{\mysignal Our Adder} &\begin{tabular}[c]{@{}c@{}}$32n+10\frac{n}{r}-7\cdot(n-1)\pmod  r $\\$ -14\omega(\frac{n}{r})-14\log n +14\log r -11$\end{tabular}
&\begin{tabular}[c]{@{}c@{}}$6 \log n+4r-3\log r$\\$ -3\log 3r+3\log (r-2)+5$\end{tabular}
 &\begin{tabular}[c]{@{}c@{}}$4n-\log n+\frac{n}{r}$\\$ -\omega(\frac{n}{r})+\log r-1$\end{tabular}\\\hline
\rowcolor{cyan!6} \multicolumn{2}{|c|}{\attherate Our Adder} &\begin{tabular}[c]{@{}c@{}}$19n+26\frac{n}{r}-7\cdot(n-1)\pmod  r $\\$ -14\omega(\frac{n}{r})-14\log n +14\log r -11$\end{tabular}
&\begin{tabular}[c]{@{}c@{}}$6 \log n+2r-3\log r$\\$ -3\log 3r+\log (r-2)+3$\end{tabular}
 &\begin{tabular}[c]{@{}c@{}}$5n-\log n- \frac{n}{r}$\\$ -\omega(\frac{n}{r})+\log r-1$\end{tabular}\\\hline
 
\end{tabular}}
\end{table*}
\egroup


\begin{table*}[!ht]
\caption{Comparison of the cost required by different quantum adders. 
The specific data of Figure \ref{fig: n1024_B} are shown below.}
\centering
\subtable[T-count Comparison\label{firsttable}]{\resizebox{0.63\linewidth}{!}{
\begin{tabular}{|l|c|c|c|c|c|c|c|} \hline
 \diagbox{Structure}{input size} & 16 & 32 & 64 & 128 & 256 & 512 & 1024\\\hline
\hline
\exclamation VBE RCA \cite{VBE}                           & 434  & 882  & 1778  & 3570  & 7154  & 14322  & 28658  \\\hline
\attherate VBE RCA\cite{VBE}                           & 134  & 262  & 518   & 1030  & 2054  & 4102   & 8198   \\\hline
\exclamation Cuccaro RCA \cite{Cuccaro}                   & 217  & 441  & 889   & 1785  & 3577  & 7161   & 14329  \\\hline
\attherate Cuccaro RCA\cite{Cuccaro}                   & 67   & 131  & 259   & 515   & 1027  & 2051   & 4099   \\\hline
\exclamation Draper In-place CLA \cite{Draper08}          & 819  & 1876 & 4053  & 8470  & 17367 & 35224  & 71001  \\\hline
\attherate Draper In-place CLA\cite{Draper08}          & 559  & 1306 & 2853  & 6000  & 12347 & 25094  & 50641  \\\hline 
\exclamation Takahashi Adder \cite{Takahashi08}           & 3136 & 6272 & 12544 & 25088 & 50176 & 100352 & 200704 \\\hline
\exclamation Takahashi RCA \cite{Takahashi09}            & 217  & 441  & 889   & 1785  & 3577  & 7161   & 14329  \\\hline
\attherate Takahashi RCA \cite{Takahashi09}            & 67   & 131  & 259   & 515   & 1027  & 2051   & 4099   \\\hline
\exclamation Takahashi Combine Adder  \cite{Takahashi09} & 784  & 1568 & 3136  & 6272  & 12544 & 25088  & 50176  \\\hline
\attherate Gidney RCA \cite{Craig_Gidney_2017}          & 60   & 124  & 252   & 508   & 1020  & 2044   & 4092   \\\hline 
\rowcolor{green!6} \exclamation Our Adder                                   & 539  & 1281 & 2954  & 6503  & 13454 & 27132  & 54607  \\\hline
\rowcolor{yellow!6}  \mysignal Our Adder                                 & 357  & 991  & 2364  & 5323  & 11283 & 23149  & 46683  \\\hline
\rowcolor{cyan!6} \attherate Our Adder                                   & 178  & 471  & 1108  & 2497  & 5317  & 10876  & 21923  \\\hline
\end{tabular}
}}
 
\qquad
 
\subtable[T-depth Comparison\label{secondtable}]{    
\resizebox{0.63\linewidth}{!}{
\begin{tabular}{|l|c|c|c|c|c|c|c|} \hline
 \diagbox{Structure}{input size} & 16 & 32 & 64 & 128 & 256 & 512 & 1024\\\hline
\hline
\exclamation VBE RCA \cite{VBE}                           & 186 & 378 & 762 & 1530 & 3066 & 6138 & 12282 \\\hline
\attherate VBE RCA  \cite{VBE}                           & 52  & 100 & 196 & 388  & 772  & 1540 & 3076  \\\hline
\exclamation Cuccaro RCA \cite{Cuccaro}                   & 93  & 189 & 381 & 765  & 1533 & 3069 & 6141  \\\hline
\attherate Cuccaro RCA  \cite{Cuccaro}                   & 18  & 34  & 66  & 130  & 258  & 514  & 1026  \\\hline
\exclamation Draper In-place CLA \cite{Draper08}          & 57  & 69  & 81  & 93   & 105  & 117  & 129   \\\hline
\attherate Draper In-place CLA  \cite{Draper08}          & 48  & 60  & 72  & 84   & 96   & 108  & 120    \\\hline 
\exclamation Takahashi Adder \cite{Takahashi08}           & 360 & 450 & 540 & 630  & 720  & 810  & 900   \\\hline
\exclamation Takahashi RCA  \cite{Takahashi09}            & 93  & 189 & 381 & 765  & 1533 & 3069 & 6141  \\\hline
\attherate Takahashi RCA   \cite{Takahashi09}            & 19  & 35  & 67  & 131  & 259  & 515  & 1027  \\\hline
\exclamation Takahashi Combine Adder   \cite{Takahashi09} & 216 & 270 & 324 & 378  & 432  & 486  & 540   \\\hline
\attherate Gidney RCA  \cite{Craig_Gidney_2017}        & 16  & 32  & 64  & 128  & 256  & 512  & 1024  \\\hline 
\rowcolor{green!6} \exclamation Our Adder  &59	&71	&83	&95	&107	&119	&131 \\\hline
 \rowcolor{yellow!6} \mysignal Our Adder &56	&68	&80	&92	&104	&116	&128   \\\hline
\rowcolor{cyan!6} \attherate Our Adder       &25	&43	&55	&67	&79	&91	&103   \\\hline
\end{tabular}
}}
\qquad
\subtable[QC Comparison\label{thirdtable}]{   
\resizebox{0.63\linewidth}{!}{
\begin{tabular}{|l|c|c|c|c|c|c|c|} \hline
 \diagbox{Structure}{input size} & 16 & 32 & 64 & 128 & 256 & 512 & 1024\\\hline
\hline
\exclamation VBE RCA \cite{VBE}                           & 49 & 97  & 193 & 385 & 769  & 1537 & 3073 \\\hline
\attherate VBE RCA  \cite{VBE}                           & 49 & 97  & 193 & 385 & 769  & 1537 & 3073 \\\hline
\exclamation Cuccaro RCA \cite{Cuccaro}                   & 34 & 66  & 130 & 258 & 514  & 1026 & 2050 \\\hline
\attherate Cuccaro RCA  \cite{Cuccaro}                   & 34 & 66  & 130 & 258 & 514  & 1026 & 2050 \\\hline
\exclamation Draper In-place CLA \cite{Draper08}          & 59 & 122 & 249 & 504 & 1015 & 2038 & 4085 \\\hline
\attherate Draper In-place CLA  \cite{Draper08}          & 59 & 122 & 249 & 504 & 1015 & 2038 & 4085 \\\hline
\exclamation Takahashi Adder \cite{Takahashi08}           & 44 & 83  & 160 & 310 & 608  & 1194 & 2355 \\\hline
\exclamation Takahashi RCA  \cite{Takahashi09}            & 33 & 65  & 129 & 257 & 513  & 1025 & 2049 \\\hline
\attherate Takahashi RCA   \cite{Takahashi09}            & 33 & 65  & 129 & 257 & 513  & 1025 & 2049 \\\hline
\exclamation Takahashi Combine Adder   \cite{Takahashi09} & 44 & 83  & 160 & 310 & 608  & 1194 & 2355 \\\hline
\attherate Gidney RCA  \cite{Craig_Gidney_2017}        & 47 & 95  & 191 & 383 & 767  & 1535 & 3071 \\\hline  
\rowcolor{green!6} \exclamation Our Adder   & 54 & 116 & 243 & 503 & 1025 & 2067 & 4150 \\\hline
\rowcolor{yellow!6}  \mysignal Our Adder& 54 & 116 & 243 & 503 & 1025 & 2067 & 4150 \\\hline
\rowcolor{cyan!6} \attherate Our Adder    & 60 & 134 & 287 & 584 & 1181 & 2373 & 4762 \\\hline

\end{tabular}
}}

\end{table*}

\begin{table*}[!ht]
\caption{Compare the cost required by Draper's out-of-place CLAs and the higher  radix adders. 
The data of the comparison between Draper's out-of-place adder and our simplified adder are shown below.}
\centering
\subtable[T-count Comparison\label{firsttable_Out-of-place}]{
{
\begin{tabular}{|l|c|c|c|c|c|c|c|} \hline
 \diagbox{Structure}{Input size} & 16 & 32 & 64 & 128 & 256 & 512 & 1024\\\hline
\hline
\exclamation Draper Out-of-place CLA \cite{Draper08}      & 448  & 987  & 2086  & 4305  & 8764  & 17703  & 35602  \\\hline
\attherate Draper Out-of-place CLA  \cite{Draper08}      & 338  & 727  & 1516  & 3105  & 6294  & 12683  & 25472  \\\hline
\rowcolor{green!6} \exclamation Our Adder   &308	&644	&1316	&2660	&5348	&10724	&21476\\\hline
 \rowcolor{yellow!6}  \mysignal Our Adder &266	&554	&1130	&2282	&4586	&9194	&18410\\\hline
\rowcolor{cyan!6} \attherate Our Adder                                     &178	&362	&730	&1466	&2938	&5882	&11770  \\\hline
\end{tabular}
}}

\subtable[T-depth Comparison\label{secondtable_Out-of-place}]{
{
\begin{tabular}{|l|c|c|c|c|c|c|c|} \hline
 \diagbox{Structure}{Input size} & 16 & 32 & 64 & 128 & 256 & 512 & 1024\\\hline
\hline
\exclamation Draper Out-of-place CLA \cite{Draper08}      & 30  & 36  & 42  & 48   & 54   & 60   & 66    \\\hline
\attherate Draper Out-of-place CLA  \cite{Draper08}      & 25  & 31  & 37  & 43   & 49   & 55   & 61    \\\hline
\rowcolor{green!6} \exclamation Our Adder                                &35&	41&	47&	53&	59&	65&	71\\\hline
 \rowcolor{yellow!6}  \mysignal Our Adder                          &38&	44&	50&	56&	62&	68&	74\\\hline
\rowcolor{cyan!6} \attherate Our Adder                                 &28&	34&	40&	46&	52&	58&	64 \\\hline
\end{tabular}
}}

\subtable[QC Comparison\label{Thirdtable_Out-of-place}]
{
{
\begin{tabular}{|l|c|c|c|c|c|c|c|} \hline
 \diagbox{Structure}{Input size} & 16 & 32 & 64 & 128 & 256 & 512 & 1024\\\hline
\hline
\exclamation Draper Out-of-place CLA \cite{Draper08}      & 60 & 123 & 250 & 505 & 1016 & 2039 & 4086 \\\hline
\attherate Draper Out-of-place CLA  \cite{Draper08}      & 60 & 123 & 250 & 505 & 1016 & 2039 & 4086 \\\hline
\rowcolor{green!6} \exclamation Our Adder   & 54 & 116 & 243 & 503 & 1025 & 2067 & 4150 \\\hline
 \rowcolor{yellow!6}  \mysignal Our Adder     & 54 & 116 & 243 & 503 & 1025 & 2067 & 4150 \\\hline
\rowcolor{cyan!6} \attherate Our Adder  & 60 & 134 & 287 & 584 & 1181 & 2373 & 4762 \\\hline
\end{tabular}
}}
\end{table*}